\documentclass[useAMS,usenatbib]{mn2e}
\usepackage{graphicx,lscape}
\usepackage{amssymb}
\usepackage{rotating}

\def\msun{M_\odot}
\def\fs{f_{\rm S}}
\def\fsst{f_{\rm S}^{\rm{st}}}

\def\Izero{{I_{\rm 0}}}
\def\I0st{{I_{\rm 0}^{\rm{st}}}}
\def\V0{V_{\rm 0}}

\def\tE{t_{\rm E}}
\def\tEgeo{t_{\rm E}^{\rm{geo}}}
\def\tEhelio{t_{\rm E}^{\rm{helio}}}
\def\te{t_{\rm E}}
\def\t0{t_{\rm 0}}
\def\uzero{u_{\rm 0}}
\def\u0{u_{\rm 0}}

\def\chidofpi{{\chi}^2_{\rm{\pi}}/\rm{N_{dof}}}

\def\tEst{t_{\rm{E}}^{\rm{st}}}
\def\tEpi{t_{\rm{E}}^{\pi}}
\def\piE{\pi_{\rm{E}}}
\def\piEvec{\vec{\pi_{E}}}
\def\piEE{\pi_{\rm{EE}}}
\def\piEN{\pi_{\rm{EN}}}
\def\murel{\mu_{\rm{rel}}}
\def\murelvec{\vec{\mu_{\rm{rel}}}}

\def\deltachi{\Delta{\chi}^2}
\def\thetaE{\theta_{\rm{E}}}
\def\pirel{\pi_{\rm{rel}}}
\def\DL{D_{\rm{L}}}
\def\DS{D_{\rm{S}}}

\def\tzeropar{t_{\rm 0}^{\rm{par}}}

\newcommand{\eg}{{e.g.},\,}
\newcommand{\ie}{{i.e.},\,}
\newcommand{\apj}{{Astrophysical Journal}}
\newcommand{\apjl}{{Astrophysical Journal Letters}}
\newcommand{\apjs}{{Astrophysical Journal Supplement Series}}
\newcommand{\araa}{{ARA\&A}}
\newcommand{\aap}{{Astronomy \& Astrophysics}}
\newcommand{\mnras}{{MNRAS}}

\newcommand{\actaa}{{Acta Astronomica}}
\newcommand{\memsai}{{Memorie della Societa Astronomica Italiana}}
\newcommand{\pasp}{{PASP}}
\newcommand{\nat}{{Nature}}

\title[BH, NS and WD Candidates from Microlensing]
{Black Hole, Neutron Star and White Dwarf Candidates from Microlensing with OGLE-III\thanks{Based on
 observations obtained with the 1.3~m Warsaw telescope at the Las Campanas Observatory of the Carnegie Institution for Science.}}
\author[{\L}. Wyrzykowski et al.]
{{\L}. Wyrzykowski$^{1}$\thanks{email: lw@astrouw.edu.pl, name
 pronunciation: {\it Woocash Vizhikovsky}}, 
 Z. Kostrzewa-Rutkowska$^{1}$,
 J. Skowron$^{1}$, \newauthor
K. A. Rybicki$^1$, P. Mr{\'o}z$^1$, S. Koz{\l}owski$^{1}$, A. Udalski$^1$,  \newauthor
M. K. Szyma{\'n}ski$^1$, G. Pietrzy{\'n}ski$^{1}$, I. Soszy{\'n}ski$^1$, K. Ulaczyk$^{1,2}$,  \newauthor
P. Pietrukowicz$^{1}$, R. Poleski$^{1,3}$, M. Pawlak$^1$, K. I{\l}kiewicz$^{1}$ 
and N. J. Rattenbury$^4$\\
$^1$ Warsaw University Astronomical Observatory, Al.~Ujazdowskie~4, 00-478~Warszawa, Poland \\
$^2$ Department of Physics, University of Warwick, Coventry CV4 7AL, UK\\
$^3$ Department of Astronomy, Ohio State University, 140 W. 18th Ave., Columbus, OH 43210, USA\\
$^4$ Department of Physics, University of Auckland, Private Bag 92019, Auckland, New Zealand
}

\begin{document}

\date{Accepted  ... Received  ...}

\pagerange{\pageref{firstpage}--\pageref{lastpage}} \pubyear{2016}

\maketitle

\label{firstpage}

\begin{abstract}
Most stellar remnants so far have been found in binary systems, where they interact with matter from their companions.
Isolated neutron stars and black holes are difficult to find as they are dark, yet they are predicted to exist in our Galaxy in vast numbers. 

We explored the OGLE-III database of 150 million objects observed in years 2001-2009 and found 59 microlensing events exhibiting a parallax effect due to the Earth's motion around the Sun. Combining parallax and brightness measurements from microlensing light curves with expected proper motions in the Milky Way, we identified 13 microlensing events which are consistent with having a white dwarf, neutron star or a black hole lens and we estimated their masses and distances. 
The most massive of our black hole candidates has 9.3 $\msun$ and is at a distance of 2.4 kpc.
The distribution of masses of our candidates indicates a continuum in mass distribution with no mass gap between neutron stars and black holes. 
We also present predictions on how such events will be observed by the astrometric Gaia mission.
\end{abstract}


\section{Introduction}

Dark stellar remnants, namely neutron stars (NS) and black holes (BH), are difficult to study as they are generally very hard to find.
Yet, they are important ingredients in our understanding of stellar evolution, mass distribution, galaxy evolution and structure, dark matter balance, etc.

Both NS and BH are typically being discovered in binary systems thanks to interactions with a companion star (\eg \citealt{Ziolkowski2010}, \citealt{2014Natur.514..202B}). 
Single neutron stars can also be found as radio pulsars when their alignment is fortunate enough so that their radio beams point toward us (\eg \citealt{2008A&A...482..617P}, \citealt{2012PPNL....9..733P}), as well as in gamma rays (\eg \citealt{2010ApJS..187..460A}).
Masses of NS and BH have been measured only in a small few dozen cases and their mass functions has been estimated (\eg \citealt{Kiziltan2013}, \citealt{Ozel2010}, \citealt{KochanekBH2015}).
The observational data collected so far indicate a clear gap between neutron stars (with masses up to 2 $\msun$) and black holes (masses from 6 $\msun$), \eg \citealt{Bailyn1998}. 
Interestingly, the theoretical predictions of the end products of stellar evolution can not reproduce the gap and rather suggest a continuum in mass distribution of remnants \citep{FryerKalogera2001}. 
This puzzle remains unresolved and the larger new sample of isolated NS and BH could unveil the answer about the mass gap.

Gravitational microlensing is the only method capable of finding isolated stellar-mass remnants and deriving their masses and the mass function (\eg \citealt{Paczynski1986}, \citealt{Dai2015}), as it does not require light emitted by the lens to detect it \citep{Paczynski2003}.
\cite{Gould2000a} estimated that about 4 per cent of lenses towards the Galactic Bulge should be due to dark remnants: neutron stars ($\sim$3 per cent) and black holes ($\sim$ 1 per cent). There should also be about 17 per cent of luminous, yet still very faint, white dwarf lenses.
More recently, \cite{Oslowski2008} simulated the BH and NS microlensing events in the Milky Way and estimated that current microlensing surveys should expect from 4 to 10 such events every year, depending on the model. 
Currently, OGLE-IV \citep{Udalski2015} detects more than 2000 microlensing events every year, therefore there should be at least a dozen of lensing black holes among them, yet none were conclusively identified as such.
The primary reason for this is that the vast majority of microlensing events are standard single point lens events (\eg \citealt{Paczynski1996}, \citealt{Wyrzykowski2015}) and such events do not allow for a determination of lens mass and distance. Additional effects and observations are required in order to break the degeneracies and to compute the mass of the lens. 
According to \cite{Gould2000b}, to measure the mass of the lens only two additional observables are needed, $\thetaE$ and $\piE$:

\begin{equation}
M=\frac{\theta_\mathrm{E}}{\kappa \pi_\mathrm{E}}
\end{equation}

where $\thetaE$ is the angular Einstein radius of the lens, constant $\kappa=4G/(c^2 AU) = 8.144~\mathrm{mas/\msun}$ and $\piE$ is microlensing parallax. 
The measurement of $\thetaE$ is possible in very rare cases, when the source star disk is being resolved during a high magnification or caustic crossing event and can be used as an angular ruler (finite source effect) (\eg \citealt{Zub}, \citealt{Lee2010}).
Another way is to detect both source and lens and to measure the relative proper motion, which leads to $\thetaE$ (\eg \citealt{GouldMass2004}, \citealt{KozlowskiHST}, \citealt{Bennett2015}, \citealt{Batista2015}). 
Yet another possibility is via precise astrometric measurements of the centroid motion during an event, however, this has not yet been achieved as the displacements to be measured are of order of 1 mas and are challenging for current facilities. In the near future, however, such measurements might be possible with the Gaia mission (\eg \citealt{BelokurovEvans2002}, \citealt{Prof2012}, \citealt{Wyrzykowski2012}) and WFIRST (The Wide-Field Infrared Survey Telescope, \citealt{BennettRhie2002}, \citealt{Spergel2015},\citealt{GouldYee2014}).

Obtaining the second parameter from Eq. 1, $\piE$, is somewhat easier. Microlensing parallax is an analog of standard parallax, and relies on detecting differences in an microlensing event caused by different viewing angles. Recently, a very successful programme of observing selected microlensing events from the Spitzer satellite, separated from the Earth more than 1 AU, has yielded parallax measurements for otherwise standard events \citep{Yee2015}.

However, in the case of long timescale events, the Earth changes its position significantly in its orbit around the Sun and therefore Earth-based parallax can be measured using only ground-based observations (\citealt{Smith2005}, \citealt{Gould2004}). 
Long time scale events are also more likely to be owing to high mass lenses. 
In an event showing the effect of parallax, the observerÕs motion changes the overall shape of the microlensing light curve, which deviates from a standard Paczy{\'n}ski curve. 

Measuring $\piE$, combined with certain assumptions, \eg on the proper motions of the lens and the source, provides probability distributions for the masses and distances of lenses.  
The terrestrial parallax effect has been used to suggest that three events from the MACHO survey are candidates for isolated black-hole lenses \citep{Bennett2002}, however, none of them was ever confirmed, despite various attempts including X-ray follow-up (\eg \citealt{Maeda2005}, \citealt{Nucita2006}). 
\cite{Mao2002} also used the parallax effect to constrain the mass of the candidate black-hole lens to about 10 $\msun$ in an event from the OGLE-II survey (OGLE-1999-BUL-32), which was also the event with the longest Einstein radius crossing time found to date ($\te$=640 days).
However, X-ray follow-up only placed an upper limit on the X-ray luminosity \citep{RevnivtsevSunyaev2001}.
There was also a candidate for a binary black hole system, OGLE-2005-SMC-001, where parallax and weak binary signatures were used to conclude that the components had masses 7 $\msun$ and 3 $\msun$  \citep{DongBH}.

In this paper we present a comprehensive study of the ground-based parallax microlensing events found in the data of the OGLE-III project (section 2).
We search for new candidates for dark remnant lenses (section 3), including candidates for black-holes and derive the mass function for isolated neutron stars and black-holes (section 4). We discuss the results in section 5 and conclude in section 6.

\section{Data}
\begin{figure*}
\center
\includegraphics[width=14.5cm]{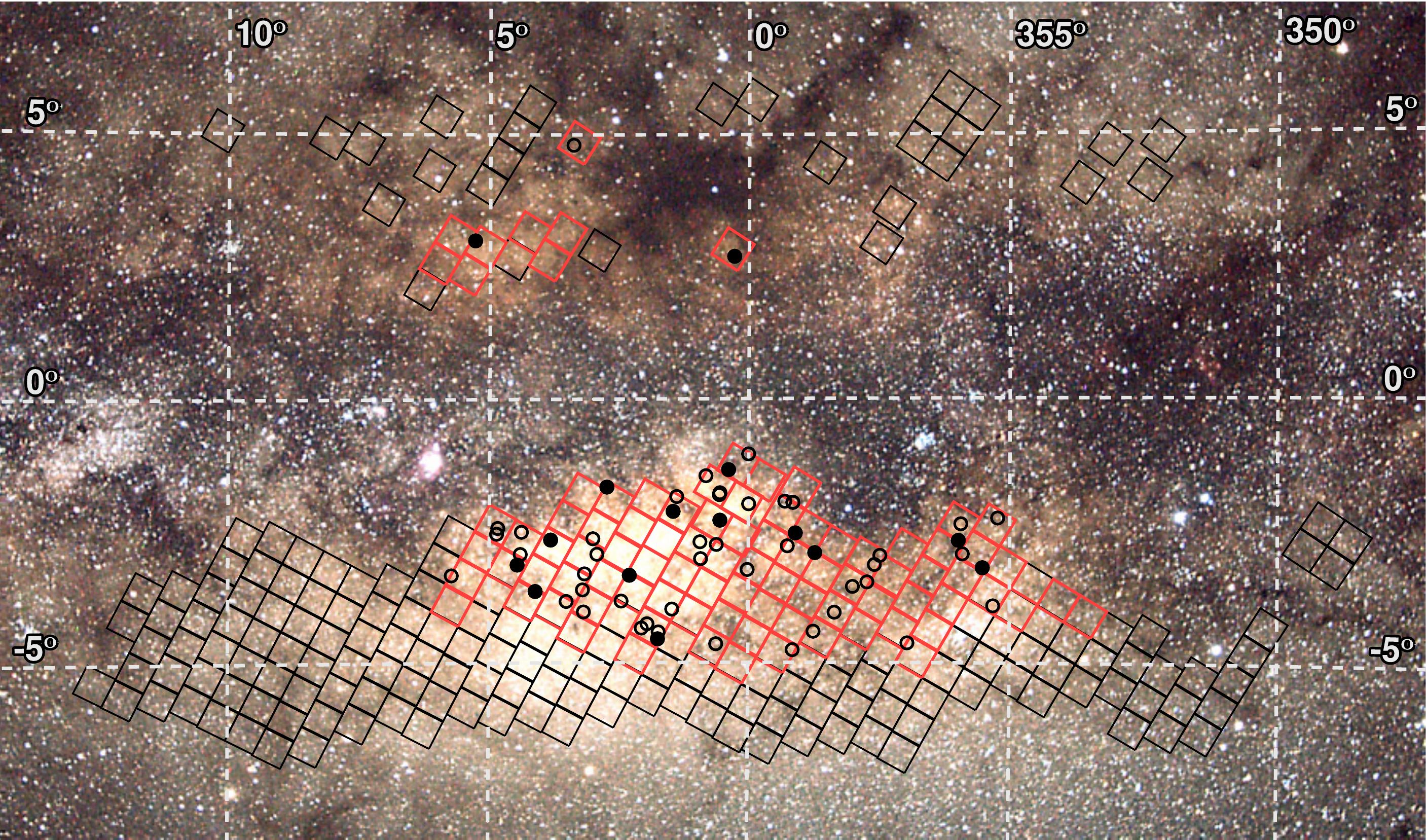}
\caption{Map of the OGLE-III bulge fields selected for search for parallax events (red squares). Black squares show other OGLE-III fields with not enough observations. Circles show 59 parallax microlensing events, while filled circles mark candidates for dark remnants lens. 
The background image was taken by Krzysztof Ulaczyk.}
\label{fig:fieldsevents}
\end{figure*}

The photometric data used here are the same as those used for our search for standard microlensing events \citep{Wyrzykowski2015}. In brief, the Optical Gravitational Lensing Experiment in its third generation (OGLE-III, \citealt{Udalski2008})  operated from 2001 until 2009 and used the 1.3m Warsaw Telescope, located at the Las Campanas Observatory, Chile, operated by the Carnegie Institution for Science.
OGLE-III observed the Galactic Bulge in 177 fields; 0.34 sq. deg each over 8 mosaic CCDs. We selected the best observed central 91 fields covering, in total, 31 sq. deg and containing about 150 million sources, each having at least 250 observations. The survey was conducted primarily in the $I$-band, calibrated to the Cousins $I$-band \citep{Szymanski2011}, and in the dense Bulge regions reached down to about 20.5 mag.
The typical sampling was, on average, once per three nights, with the exception of the most central fields, which (from 2005) were observed about three times per night, yielding up to 2500 data points for those regions.
Fig. \ref{fig:fieldsevents} shows a map of the OGLE-III fields with all microlensing events exhibiting a parallax effect marked.

$V$-band images were collected occasionally, with up to 35 data points available per light curve, providing an opportunity to derive average colours of the stars. 

The images in both bands were reduced on-the-fly soon after the observations using Difference Imaging Analysis (DIA, \citealt{Wozniak2000}), which allowed for real-time discoveries of on-going microlensing events via OGLE's Early Warning System (EWS\footnote{http://ogle.astrouw.edu.pl/ogle3/ews/ews.html}, \citealt{Udalski2003}). At the end of OGLE-III the data were re-reduced using better quality reference images \citep{Udalski2008}. This is the primary database in which we conducted our search. 
The microlensing events discovered among the 150 million light curves in the database were then additionally re-reduced further, using more precise light centroid positions, in order to improve their photometry. All microlensing modelling was performed using this new photometry, but the search itself and the simulation of events used the main data base, in order to maintain consistency. 
The photometric error-bars were corrected using scaling coefficients derived for each field/chip (see Wyrzykowski et al. 2009). A few frames were removed from the analysis as they were taken under very poor conditions (mostly from the very first season of OGLE-III, when the survey was just starting) and obvious outlying data points were removed using a  sigma-clipping method.

\section{Search for remnant lenses}
\label{search}

We explored the vast OGLE-III database of light curves containing about 150 million sources over a number of stages in order to find microlensing events showing a parallax effect, which could be attributed to stellar remnant lenses.

\subsection{Preselection of objects}
As the first step we take the simple and quick to compute von Neumann ($\eta$) and skewness ($\gamma$) statistics, previously computed for all stars in the 91 OGLE-III fields in the search for dips in the OGLE light curves \citep{Rattenbury2014}. 
The combination of those two statistics is sensitive to temporal changes in brightness -- a negative value of skewness indicates a brightening, like in microlensing events. We then visually selected a couple of dozen long-lasting microlensing events (found during the search for standard events in \citealt{Wyrzykowski2015}) and used them to derive the boundary in the $\eta$ and $\gamma$ parameter space.

The cut was able to limit a large sample of stars to only those which exhibited a long-term smooth variations (16,060). Among those we removed stars which were identified as long period variable stars in \cite{Soszynski2013} and ghost events \cite{Wyrzykowski2015}, which often were multiplied occurrences of the same object. We were then left with just 3845 light curves.
Those were then fitted with the standard Paczy{\'n}ski microlensing model for a single lens, with a blending parameter fixed at a set of values ranging from 1 to 0.1, to avoid models not converging numerically. At this stage we used a simple and quick to compute Levenberg-Marquardt Least-Square (LSQ) minimization algorithm in order to filter out genuine microlensing events with parallax from other outburst-like and variable objects. 
The parameters of the standard model were the following: $\t0^{\rm{st}}$ - time of the closest lens-source approach; $\u0^{\rm{st}}$ - minimum impact parameter; $\tEst$ - Einstein radius crossing time (the eventÕs time-scale); $\I0st$ - baseline magnitude; $\fsst$ - blending parameter (defined as the ratio of the source flux contribution to the total baseline flux). 

In the next step, the parameters of the standard model were used to seed the search for a parallax microlensing model (Gould 2004). 
The parameters of the parallax model, computed in the geocentric frame, included five parameters as in the standard model (but indexed with $\pi$ in order to distinguish from their standard model equivalents) and two additional parameters: $\piEN$ and $\piEE$, the North and East components, respectively, of the parallax vector, $\piEvec$. The geocentric parallax model also requires a reference time ($\tzeropar$), which was chosen to be $\t0^{\rm{st}}$.

While fitting the parallax model with LSQ the blending parameter ($\fs$) was again fixed at set values from 1 to 0.1 in order to avoid non-convergence of models (but independent on $\fs$ derived for the standard model, as the parallax effect can sometimes mimic blending in the standard model).

We then computed the difference between the goodness of fit between the parallax and standard models ($\deltachi$), which was used as the main indicator for the significant parallax signal. 
We also added an empirical parameter, $ampl$, which was the amplitude of the light curve, in order to exclude very shallow and long-term variable objects, not caused by microlensing. 

We have visually inspected all 3845 objects remaining after the rejection of variable stars and ghosts and selected 59 microlensing events which were clearly exhibiting parallax-like deviation from the standard model. The visually selected sample (hereafter referred to as the {\it golden sample}) allowed us to construct a set of preliminary cuts on various parameters of the fitted models in order to narrow down and automatise the selection.
The series of cuts used microlensing parameters from both standard and parallax LSQ models, along with quantities like the amplitude of variation and other statistics. 
The entire sequence of the cuts applied is described in Table \ref{tab:cuts}. 
The results of the cuts is a set of 217 objects, which includes all 59 golden sample parallax events. 

\begin{table*}
\begin{tiny}
\centering
\caption{Cuts applied to the entire OGLE-III database in order to detect events with a clear parallax signal. }
\label{tab:cuts}
\begin{tabular}{llll}
\hline
Cut number & Formula & Description & N objects remaining  \\
\hline
0 & 91 fields, at least 250 datapoints & initial database & 150$\times 10^6$ \\
 & & & \\
1. & $\gamma<0$ \& $\log(-\gamma)>-1.25(\log(1/\eta)+0.75$ & up-wards and long-term deviations & 16060 \\
 & & & \\
2. & NOT var \& NOT ghost & removed known variable stars and ghost events & 3845\\
 & & & \\
3. & $ 2050<JD(\t0)-2450000<5000$ & maximum within the data range & 3645 \\
 & & & \\
4. & $\deltachi>50$ \& $\deltachi>75(\chidofpi -3)+50$ & parallax model significantly better than standard & 679 \\
 & & & \\
5. & $\tEst>20$ \& $\tEst<650$ \& $\tEst>13(\chidofpi-3)+5$ & non-parallax time-scale (duration of the event in days) above thresholds & 389 \\
 & & & \\
6. & $\rm{ampl}>0.16$ mag & significant amplitude of variation & 318 \\
 & & & \\
7. & $\tEpi>50$ d & parallax model time-scale long enough & 245 \\
 & & & \\
8. & $\chidofpi<10$ & parallax model well fit & 217 \\
 & & & \\
(9.) & & visual inspection/Random Forest classifier & 59 \\
\hline
\end{tabular}
\end{tiny}
\end{table*}

\subsubsection{Random Forest classifier}
The cuts on parameters described above were not ideal as they were simple linear cuts through the parameter space and the selected objects still required a visual inspection. 
In order to prepare a fully automated search pipeline, it was necessary to replace the visual inspection with more complicated selection criteria, which can be found using Machine Learning algorithms. We therefore substituted the human with a similarly skilled Random Forest (RF) classifier (\citealt{Breiman2001}, \citealt{Wyrzykowski2015}). 

The training of the RF was performed on the entire sample of 217 events which remained after the cuts, using their parameters, \eg microlensing fit parameters of standard and parallax model, $\eta$, $\gamma$, $\Delta\chi^2$ between the goodness of fit between parallax and standard models. The cross-validation of that training (as run in the Weka\footnote{http://www.cs.waikato.ac.nz/ml/weka/} package) yielded about 81 per cent correct classifications, however, the entire golden sample was retrieved without any losses when run on the whole sample. This guarantees that the classifier did not exclude good events, but performed well on removing bad cases, \ie with high completeness. The classifier would be used for simulated events only, hence its primary role was not to detect and remove non-microlensing events, but to recognise microlensing events with the  parallax effect. 

Table \ref{tab:names} lists all 59 golden sample parallax microlensing events, their names, OGLE database IDs, coordinates and length, defined as the time from detectable microlensing deviation computed from the parallax model (\ie by 1$\sigma$ from the baseline).

\subsubsection{MCMC Modelling}
Each of the selected events from the golden sample was modelled with the Markov chain Monte Carlo (MCMC) method. Full MCMC runs were made across on an extended range of parameters allowing all degenerate solutions for microlensing parallax model to be found (\eg \citealt{Gould2004}, \citealt{Smith2005}) as well as obtaining probability distributions for each of the parameters of the model.
To find the microlensing parameters and their 1-$\sigma$ limits we use the Affine Invariant MCMC Ensemble sampler - \textit{emcee} (\cite{EMCEE}).
Note, in the MCMC parallax microlensing models we allowed for a small amount of negative ``third light'', so called, negative blending. 
In our definition of blending, the negative blending occurrs when $fs>1$ \citep{Smith2007}. 
Small amount of negative blending is an indication of possible over-estimation of the background level on the reference image, locally influenced by numerous unresolved stars (see also \citealt{Wyrzykowski2015}). We also tested the convergence of the MCMC chains for unconstrained $\fs$ and found that some of the solutions were nonphysical ($\fs>1$), hence those were removed. 

We denote the microlensing parallax model parameters obtained from the MCMC modelling with no additional indices, in order to distinguish them from the parameters from the quick LSQ modelling performed earlier in the search pipeline. 

The parameters from the MCMC modelling for all 59 events are available on-line\footnote{http://ogle.astrouw.edu.pl} and in the on-line version of the paper. 
Table \ref{tab:mcmc-rem} shows the microlensing parallax model parameters for 13 events selected as the most probable candidates for having dark stellar remnants as lenses (see next section).

\begin{table*}
\begin{scriptsize}
\centering
\caption{Parallax microlensing events found in OGLE-III. Columns show the unique name of the lens, OGLE database id, right-ascension, declination, OGLE-III EWS match and length of the light curve.}
\label{tab:names}
\begin{tabular}{llcccc}
\hline
Lens name & OGLE DB ID & RA$_\mathrm{J2000}$ & DEC$_\mathrm{J2000}$ & EWS & length$^*$ \\
OGLE3-ULENS- & [field.chip.starno] & [h:m:s] & [d:m:s] & OGLE- & [days] \\
\hline
\hline
PAR-01 & BLG196.1.118200 & 18:04:48.79 & -29:40:31.1 & --- & 2840.8\\
PAR-02 & BLG205.3.159237 & 17:57:23.14 & -28:46:32.0 & 2006-BLG-095 & 2306.4\\
PAR-03 & BLG242.7.31173 & 18:08:0.42 & -26:39:15.4 & --- & 2240.3\\
PAR-04 & BLG342.5.73806 & 17:46:25.67 & -22:51:37.3 & --- & 2225.4\\
PAR-05 & BLG138.1.192949 & 17:46:35.45 & -33:46:19.8 & 2004-BLG-361 & 1947.5\\
PAR-06 & BLG225.8.163880 & 18:03:52.40 & -27:54:1.3 & 2006-BLG-393 & 1566.3\\
PAR-07 & BLG102.7.44461 & 17:55:59.44 & -29:38:28.8 & 2005-BLG-474 & 1244.8\\
PAR-08 & BLG130.1.122103 & 17:47:34.77 & -34:25:50.3 & 2005-BLG-351 & 1223.2\\
PAR-09 & BLG172.8.614 & 17:54:11.82 & -31:31:57.9 & --- & 1141.5\\
PAR-10 & BLG156.8.68730 & 17:54:4.49 & -32:40:38.2 & --- & 1044.2\\
PAR-11 & BLG197.3.171180 & 18:07:33.38 & -29:20:16.2 & 2003-BLG-263 & 1030.3\\
PAR-12 & BLG129.6.16369 & 17:42:58.78 & -34:11:13.2 & 2005-BLG-059 & 991.0\\
PAR-13 & BLG333.2.53483 & 17:35:55.95 & -27:16:2.1 & 2002-BLG-061 & 972.9\\
PAR-14 & BLG101.7.52163 & 17:53:28.47 & -29:57:44.7 & 2008-BLG-318 & 970.6\\
PAR-15 & BLG223.5.115009 & 17:58:27.13 & -27:26:0.9 & 2008-BLG-096 & 935.2\\
PAR-16 & BLG172.5.160833 & 17:54:55.05 & -31:00:40.5 & 2003-BLG-032 & 928.9\\
PAR-17 & BLG249.2.92350 & 18:06:6.30 & -25:59:12.8 & 2008-BLG-157 & 927.8\\
PAR-18 & BLG205.5.12205 & 17:56:7.08 & -28:41:37.3 & --- & 848.8\\
PAR-19 & BLG194.3.95955 & 17:51:48.90 & -29:17:52.9 & 2005-BLG-372 & 826.8\\
PAR-20 & BLG197.6.98763 & 18:06:59.34 & -29:23:39.7 & 2008-BLG-098 & 800.9\\
PAR-21 & BLG122.1.184151 & 17:50:4.59 & -34:58:15.3 & 2006-BLG-005 & 790.3\\
PAR-22 & BLG158.5.38090 & 18:00:20.30 & -32:15:11.3 & 2002-BLG-334 & 755.1\\
PAR-23 & BLG224.3.40806 & 18:02:53.14 & -27:41:15.7 & --- & 747.5\\
PAR-24 & BLG236.7.59423 & 18:09:13.92 & -27:12:12.6 & --- & 747.2\\
PAR-25 & BLG251.4.149142 & 18:11:29.35 & -25:38:24.0 & 2005-BLG-036 & 747.1\\
PAR-26 & BLG130.4.125656 & 17:47:27.03 & -33:58:8.4 & 2006-BLG-020 & 745.9\\
PAR-27 & BLG234.6.218982 & 18:04:45.71 & -26:59:15.3 & 2005-BLG-086 & 735.4\\
PAR-28 & BLG197.1.69448 & 18:07:41.45 & -29:42:39.1 & 2005-BLG-020 & 716.7\\
PAR-29 & BLG218.3.139976 & 18:08:46.04 & -28:12:34.6 & 2006-BLG-031 & 680.6\\
PAR-30 & BLG104.1.148712 & 17:59:44.30 & -29:41:7.3 & 2008-BLG-545 & 677.8\\
PAR-31 & BLG167.8.67944 & 18:02:41.84 & -32:04:39.6 & --- & 672.0\\
PAR-32 & BLG195.2.8006 & 17:53:55.65 & -29:22:41.1 & 2008-BLG-223 & 667.1\\
PAR-33 & BLG208.5.131806 & 18:04:4.02 & -28:38:38.8 & 2006-BLG-251 & 661.0\\
PAR-34 & BLG180.6.142446 & 17:51:38.96 & -30:32:16.9 & --- & 604.5\\
PAR-35 & BLG354.3.45010 & 17:35:25.94 & -23:31:0.7 & 2003-BLG-297 & 583.9\\
PAR-36 & BLG208.3.222797 & 18:06:21.76 & -28:44:46.3 & --- & 555.0\\
PAR-37 & BLG195.5.141274 & 17:53:16.57 & -28:58:49.8 & --- & 552.4\\
PAR-38 & BLG155.8.123452 & 17:51:25.67 & -32:38:6.1 & --- & 510.9\\
PAR-39 & BLG171.4.87121 & 17:53:34.61 & -31:01:12.3 & 2008-BLG-013 & 505.2\\
PAR-40 & BLG157.3.1531 & 17:57:54.73 & -32:25:31.9 & 2005-BLG-454 & 492.5\\
PAR-41 & BLG226.1.1251 & 18:07:8.81 & -27:59:22.5 & 2003-BLG-459 & 467.3\\
PAR-42 & BLG134.5.193547 & 17:56:59.79 & -33:55:15.6 & 2005-BLG-061 & 463.1\\
PAR-43 & BLG182.3.104094 & 17:58:30.91 & -30:34:7.6 & 2009-BLG-049 & 459.7\\
PAR-44 & BLG180.7.173170 & 17:51:21.97 & -30:41:1.2 & 2007-BLG-282/2007-BLG-303 & 452.8\\
PAR-45 & BLG226.8.198790 & 18:05:51.80 & -27:52:23.3 & 2006-BLG-015 & 428.5\\
PAR-46 & BLG156.2.110020 & 17:55:5.17 & -32:28:51.4 & 2006-BLG-023 & 427.9\\
PAR-47 & BLG179.2.183955 & 17:51:21.96 & -30:41:1.2 & 2007-BLG-282/2007-BLG-303 & 417.3\\
PAR-48 & BLG197.1.101151 & 18:07:8.11 & -29:39:34.0 & 2008-BLG-375 & 410.8\\
PAR-49 & BLG185.7.91655 & 18:05:34.74 & -30:44:20.2 & 2007-BLG-035 & 398.1\\
PAR-50 & BLG227.7.16177 & 18:08:41.73 & -27:49:14.9 & --- & 338.3\\
PAR-51 & BLG249.2.60315 & 18:06:34.67 & -26:01:16.0 & 2003-BLG-175 & 333.6\\
PAR-52 & BLG241.2.105532 & 18:07:5.68 & -26:36:33.1 & 2008-BLG-038 & 325.6\\
PAR-53 & BLG195.2.18171 & 17:54:5.53 & -29:23:11.0 & 2004-BLG-070 & 323.3\\
PAR-54 & BLG104.7.157693 & 17:58:29.56 & -29:30:54.3 & 2006-BLG-366 & 322.4\\
PAR-55 & BLG138.7.140195 & 17:45:11.29 & -33:38:55.2 & 2006-BLG-061 & 318.2\\
PAR-56 & BLG155.1.109582 & 17:52:23.34 & -32:37:57.9 & --- & 296.5\\
PAR-57 & BLG241.6.137439 & 18:05:24.43 & -26:25:18.8 & 2007-BLG-349 & 261.5\\
PAR-58 & BLG188.5.139006 & 17:57:58.60 & -29:48:48.4 & --- & 216.7\\
PAR-59 & BLG194.7.63636 & 17:49:44.34 & -29:28:34.3 & --- & 115.4\\
\hline
\hline
\end{tabular}
\newline
$^*$ Length is the duration of the entire event, defined as the time between measurable microlensing deviation, depending on the photometric error-bar (see text).
\end{scriptsize}
\end{table*}

\begin{table*}
\caption{Microlensing parallax model parameters as obtained from MCMC modelling for 13 dark remnant candidate events. Multiple entries per event indicate different degenerate solutions found. The full table with the parameters for all 59 events is available online.}
\begin{sideways}
\begin{tiny}
\label{tab:mcmc-rem} 
\begin{tabular}{lccccccccccc}
\hline
Lens name & $\tzeropar$ & $\t0$ & $\tEgeo$ & $u_{\mathrm{0}}$ & $\piEN$ & $\piEE$ & $I_{\rm{0}}$ & $\fs$ & $\tEhelio$ & $\chi^2/ndof$ & ndof \\
OGLE3-ULENS- & [days] & [days] & [days] &  & &  & [mag] & & [days] & & \\
\hline
\hline
PAR-02 & 4018.6 & $4064.2^{+0.9}_{-0.9}$ & $317.9^{+8.2}_{-8.0}$ & $0.65182^{+0.02603}_{-0.02486}$ & $0.033^{+0.001}_{-0.001}$ & $-0.051^{+0.002}_{-0.002}$ & $15.453^{+0.001}_{-0.001}$ & $0.64^{+0.04}_{-0.04}$ & $296.1^{+7.6}_{-7.4}$ & 1.14 & 852\\
PAR-02 & 4018.6 & $4077.1^{+0.4}_{-0.4}$ & $288.8^{+8.1}_{-6.0}$ & $-0.83380^{+0.03757}_{-0.02844}$ & $-0.033^{+0.001}_{-0.001}$ & $-0.073^{+0.004}_{-0.003}$ & $15.452^{+0.001}_{-0.000}$ & $1.01^{+0.07}_{-0.08}$ & $255.9^{+7.4}_{-5.3}$ & 1.13 & 852\\
PAR-04 & 3306.5 & $3340.3^{+0.2}_{-0.3}$ & $143.1^{+3.7}_{-2.9}$ & $0.13919^{+0.00631}_{-0.00707}$ & $0.125^{+0.003}_{-0.003}$ & $0.165^{+0.005}_{-0.006}$ & $14.536^{+0.001}_{-0.001}$ & $1.04^{+0.04}_{-0.05}$ & $184.0^{+4.0}_{-3.2}$ & 1.16 & 297\\
PAR-04 & 3306.5 & $3339.7^{+0.3}_{-0.3}$ & $143.7^{+3.6}_{-3.2}$ & $-0.14181^{+0.00718}_{-0.00668}$ & $-0.128^{+0.003}_{-0.003}$ & $0.154^{+0.005}_{-0.005}$ & $14.536^{+0.001}_{-0.001}$ & $1.03^{+0.04}_{-0.04}$ & $185.2^{+4.0}_{-3.5}$ & 1.10 & 297\\
PAR-05 & 3512.7 & $3503.5^{+0.5}_{-0.5}$ & $161.8^{+5.3}_{-3.9}$ & $0.75628^{+0.02793}_{-0.03512}$ & $0.074^{+0.002}_{-0.002}$ & $-0.041^{+0.002}_{-0.002}$ & $14.828^{+0.000}_{-0.000}$ & $1.00^{+0.07}_{-0.08}$ & $174.1^{+5.0}_{-3.7}$ & 1.47 & 754\\
PAR-05 & 3512.7 & $3503.6^{+0.4}_{-0.4}$ & $152.2^{+4.4}_{-4.2}$ & $-0.72974^{+0.03182}_{-0.03281}$ & $-0.084^{+0.003}_{-0.003}$ & $-0.060^{+0.003}_{-0.003}$ & $14.827^{+0.000}_{-0.000}$ & $0.94^{+0.08}_{-0.07}$ & $175.5^{+4.7}_{-4.5}$ & 1.59 & 754\\
PAR-07 & 3856.8 & $3859.1^{+0.1}_{-0.1}$ & $120.0^{+2.3}_{-2.2}$ & $0.41701^{+0.01335}_{-0.01339}$ & $-0.074^{+0.010}_{-0.009}$ & $-0.089^{+0.006}_{-0.006}$ & $15.760^{+0.000}_{-0.000}$ & $0.92^{+0.04}_{-0.04}$ & $137.5^{+1.8}_{-1.8}$ & 1.78 & 1021\\
PAR-07 & 3856.8 & $3859.1^{+0.1}_{-0.1}$ & $134.1^{+1.5}_{-1.5}$ & $-0.34862^{+0.00824}_{-0.00783}$ & $0.042^{+0.012}_{-0.014}$ & $-0.070^{+0.005}_{-0.005}$ & $15.760^{+0.000}_{-0.000}$ & $0.71^{+0.02}_{-0.02}$ & $147.9^{+1.4}_{-1.4}$ & 1.82 & 1021\\
PAR-08 & 3779.9 & $3745.8^{+1.6}_{-1.6}$ & $153.2^{+11.4}_{-6.9}$ & $1.02360^{+0.07136}_{-0.10225}$ & $0.173^{+0.006}_{-0.006}$ & $-0.016^{+0.007}_{-0.007}$ & $15.715^{+0.000}_{-0.000}$ & $0.91^{+0.14}_{-0.17}$ & $135.4^{+9.5}_{-5.8}$ & 1.95 & 781\\
PAR-08 & 3779.9 & $3757.1^{+0.9}_{-0.9}$ & $136.3^{+7.2}_{-5.8}$ & $-1.01386^{+0.10536}_{-0.08769}$ & $-0.236^{+0.010}_{-0.010}$ & $-0.037^{+0.016}_{-0.014}$ & $15.715^{+0.000}_{-0.000}$ & $0.86^{+0.16}_{-0.17}$ & $135.4^{+9.1}_{-6.6}$ & 1.82 & 781\\
PAR-09 & 4268.8 & $4275.6^{+0.4}_{-0.4}$ & $127.3^{+4.3}_{-3.5}$ & $0.69147^{+0.02975}_{-0.03344}$ & $-0.199^{+0.005}_{-0.005}$ & $-0.151^{+0.005}_{-0.005}$ & $16.833^{+0.001}_{-0.001}$ & $1.00^{+0.08}_{-0.08}$ & $159.1^{+5.3}_{-4.4}$ & 2.20 & 911\\
PAR-13 & 2510.6 & $2509.7^{+0.2}_{-0.2}$ & $185.3^{+16.3}_{-11.5}$ & $0.16384^{+0.02066}_{-0.02109}$ & $-0.035^{+0.052}_{-0.047}$ & $0.058^{+0.005}_{-0.004}$ & $17.526^{+0.001}_{-0.001}$ & $0.72^{+0.11}_{-0.11}$ & $171.7^{+11.8}_{-8.8}$ & 1.79 & 324\\
PAR-13 & 2510.6 & $2509.7^{+0.2}_{-0.2}$ & $181.9^{+6.7}_{-5.7}$ & $-0.16480^{+0.01257}_{-0.01247}$ & $0.060^{+0.050}_{-0.060}$ & $0.062^{+0.005}_{-0.004}$ & $17.526^{+0.001}_{-0.001}$ & $0.73^{+0.06}_{-0.06}$ & $170.6^{+7.1}_{-6.1}$ & 1.78 & 324\\
PAR-15 & 4720.6 & $4721.9^{+0.5}_{-0.4}$ & $131.5^{+5.7}_{-5.3}$ & $0.49237^{+0.03994}_{-0.03779}$ & $-0.036^{+0.016}_{-0.017}$ & $-0.140^{+0.006}_{-0.006}$ & $16.831^{+0.000}_{-0.000}$ & $0.87^{+0.11}_{-0.10}$ & $138.2^{+5.6}_{-5.3}$ & 1.18 & 797\\
PAR-15 & 4720.6 & $4721.7^{+0.4}_{-0.4}$ & $131.2^{+4.5}_{-4.4}$ & $-0.46919^{+0.03295}_{-0.03696}$ & $0.070^{+0.023}_{-0.023}$ & $-0.142^{+0.006}_{-0.007}$ & $16.831^{+0.000}_{-0.000}$ & $0.80^{+0.10}_{-0.08}$ & $140.6^{+5.2}_{-5.1}$ & 1.19 & 797\\
PAR-19 & 3776.9 & $3725.7^{+0.3}_{-0.2}$ & $87.7^{+2.8}_{-2.0}$ & $0.11757^{+0.03619}_{-0.05388}$ & $-0.069^{+0.012}_{-0.015}$ & $0.036^{+0.007}_{-0.009}$ & $15.762^{+0.000}_{-0.000}$ & $1.02^{+0.06}_{-0.08}$ & $91.4^{+2.4}_{-1.7}$ & 1.48 & 889\\
PAR-19 & 3776.9 & $3726.4^{+0.2}_{-0.2}$ & $90.2^{+2.1}_{-2.3}$ & $-0.02133^{+0.01564}_{-0.02797}$ & $-0.118^{+0.020}_{-0.023}$ & $0.033^{+0.010}_{-0.009}$ & $15.762^{+0.000}_{-0.000}$ & $0.93^{+0.05}_{-0.04}$ & $94.2^{+1.5}_{-1.6}$ & 1.49 & 889\\
PAR-27 & 3634.5 & $3628.3^{+0.3}_{-0.3}$ & $93.5^{+2.0}_{-1.9}$ & $0.46645^{+0.00739}_{-0.01256}$ & $0.245^{+0.031}_{-0.034}$ & $0.110^{+0.008}_{-0.008}$ & $16.488^{+0.001}_{-0.001}$ & $1.08^{+0.03}_{-0.04}$ & $95.2^{+2.3}_{-2.2}$ & 1.19 & 593\\
PAR-27 & 3634.5 & $3630.7^{+0.4}_{-0.4}$ & $107.2^{+7.1}_{-7.7}$ & $0.29440^{+0.04488}_{-0.03217}$ & $-0.162^{+0.076}_{-0.043}$ & $0.081^{+0.006}_{-0.007}$ & $16.487^{+0.001}_{-0.001}$ & $0.57^{+0.12}_{-0.08}$ & $104.8^{+6.1}_{-6.3}$ & 1.24 & 593\\
PAR-27 & 3634.5 & $3631.7^{+0.2}_{-0.2}$ & $119.8^{+5.2}_{-5.2}$ & $-0.23097^{+0.01358}_{-0.01579}$ & $0.336^{+0.015}_{-0.018}$ & $0.093^{+0.006}_{-0.006}$ & $16.488^{+0.001}_{-0.001}$ & $0.41^{+0.04}_{-0.03}$ & $124.2^{+5.7}_{-5.7}$ & 1.20 & 593\\
PAR-27 & 3634.5 & $3627.9^{+0.4}_{-0.4}$ & $108.0^{+4.0}_{-3.9}$ & $-0.41246^{+0.01561}_{-0.01640}$ & $-0.296^{+0.031}_{-0.029}$ & $0.132^{+0.012}_{-0.012}$ & $16.488^{+0.001}_{-0.001}$ & $0.90^{+0.05}_{-0.05}$ & $103.6^{+3.5}_{-3.4}$ & 1.19 & 593\\
PAR-28 & 3496.7 & $3491.9^{+0.2}_{-0.2}$ & $79.7^{+4.2}_{-2.6}$ & $0.49072^{+0.03109}_{-0.04304}$ & $0.152^{+0.052}_{-0.036}$ & $0.126^{+0.006}_{-0.006}$ & $15.242^{+0.000}_{-0.000}$ & $0.96^{+0.10}_{-0.12}$ & $69.8^{+2.4}_{-1.7}$ & 1.45 & 770\\
PAR-28 & 3496.7 & $3492.2^{+0.2}_{-0.2}$ & $73.7^{+0.8}_{-0.7}$ & $-0.52878^{+0.01175}_{-0.00658}$ & $-0.256^{+0.040}_{-0.041}$ & $0.113^{+0.006}_{-0.006}$ & $15.242^{+0.000}_{-0.000}$ & $1.08^{+0.02}_{-0.04}$ & $67.4^{+0.8}_{-0.8}$ & 1.46 & 770\\
PAR-33 & 4012.5 & $4008.3^{+0.1}_{-0.1}$ & $79.9^{+1.0}_{-0.9}$ & $0.32979^{+0.00650}_{-0.00729}$ & $0.210^{+0.041}_{-0.047}$ & $-0.065^{+0.009}_{-0.009}$ & $16.222^{+0.000}_{-0.000}$ & $1.06^{+0.03}_{-0.03}$ & $80.5^{+1.1}_{-1.1}$ & 3.28 & 909\\
PAR-33 & 4012.5 & $4009.5^{+0.1}_{-0.1}$ & $113.4^{+5.8}_{-6.1}$ & $0.17354^{+0.01404}_{-0.01166}$ & $-0.377^{+0.026}_{-0.019}$ & $-0.048^{+0.011}_{-0.012}$ & $16.222^{+0.000}_{-0.000}$ & $0.47^{+0.05}_{-0.04}$ & $103.9^{+4.7}_{-4.9}$ & 3.29 & 909\\
PAR-33 & 4012.5 & $4009.5^{+0.1}_{-0.1}$ & $85.8^{+1.9}_{-1.7}$ & $-0.24572^{+0.00868}_{-0.00902}$ & $0.378^{+0.028}_{-0.033}$ & $-0.049^{+0.009}_{-0.009}$ & $16.222^{+0.000}_{-0.000}$ & $0.72^{+0.03}_{-0.03}$ & $88.5^{+2.3}_{-2.2}$ & 3.28 & 909\\
PAR-33 & 4012.5 & $4007.9^{+0.1}_{-0.1}$ & $93.7^{+2.6}_{-2.6}$ & $-0.28215^{+0.00667}_{-0.00680}$ & $-0.300^{+0.036}_{-0.031}$ & $-0.062^{+0.009}_{-0.009}$ & $16.222^{+0.000}_{-0.000}$ & $0.86^{+0.03}_{-0.03}$ & $87.9^{+2.1}_{-2.0}$ & 3.27 & 909\\
PAR-39 & 4576.8 & $4577.1^{+0.0}_{-0.0}$ & $75.9^{+0.3}_{-0.3}$ & $0.02286^{+0.00012}_{-0.00014}$ & $-0.066^{+0.068}_{-0.074}$ & $-0.038^{+0.008}_{-0.009}$ & $16.318^{+0.000}_{-0.000}$ & $0.90^{+0.00}_{-0.01}$ & $78.5^{+1.0}_{-0.9}$ & 1.05 & 778\\
PAR-39 & 4576.8 & $4577.1^{+0.0}_{-0.0}$ & $75.7^{+0.5}_{-0.4}$ & $-0.02284^{+0.00014}_{-0.00012}$ & $-0.071^{+0.077}_{-0.083}$ & $-0.038^{+0.008}_{-0.009}$ & $16.318^{+0.000}_{-0.000}$ & $0.90^{+0.00}_{-0.01}$ & $78.3^{+0.9}_{-0.8}$ & 1.05 & 778\\
\hline
\hline
\end{tabular}
\end{tiny}
\end{sideways} 
\end{table*}

\subsubsection{Colour-magnitude diagram selection}

\begin{figure}
\centering
\includegraphics[width=8.5cm]{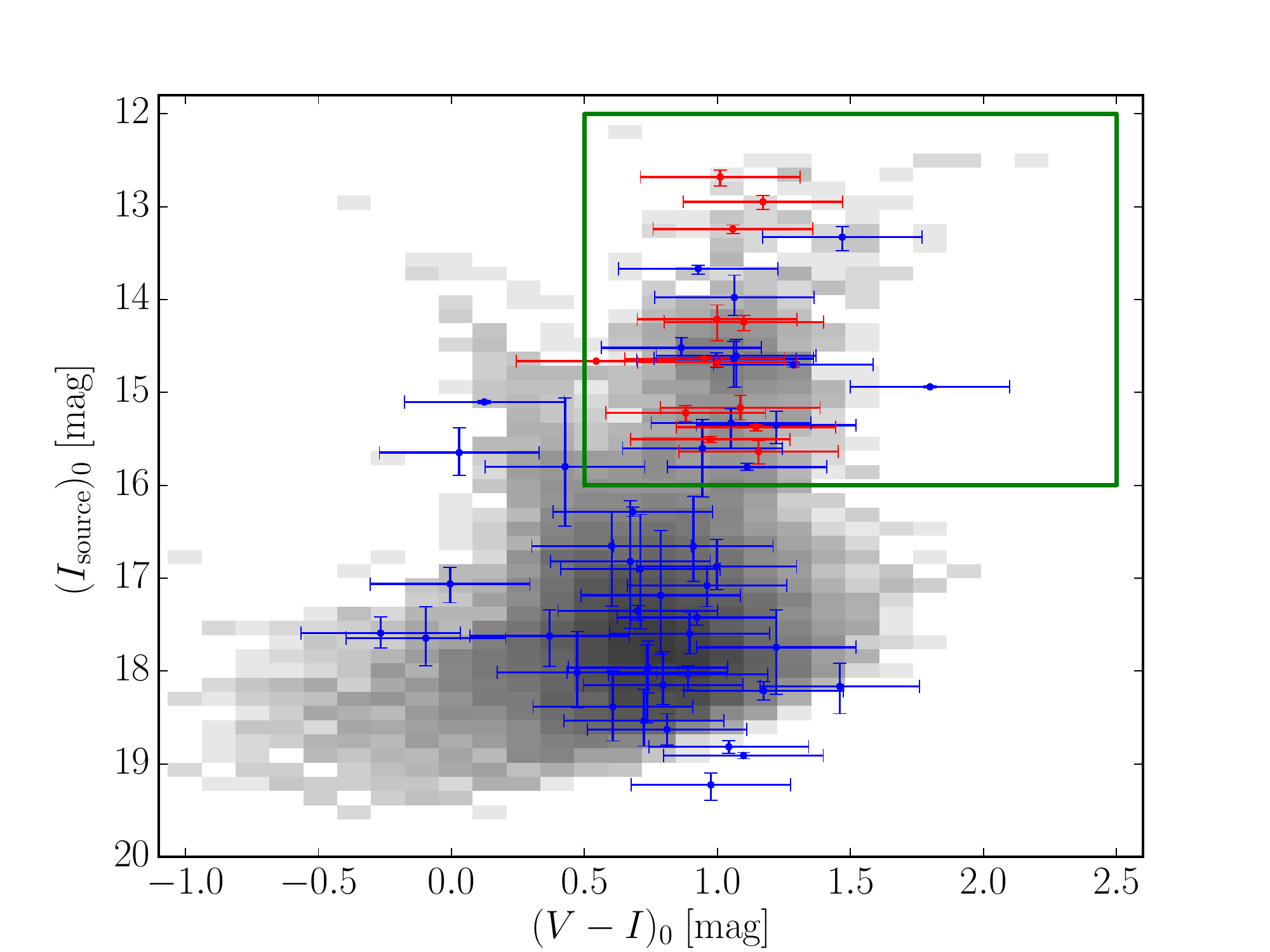} 
\caption{Extinction free colour-magnitude diagram for sources in our sample of parallax events from OGLE-III. The background is the extinction-corrected CMD of stars in the sub-field BLG205.3. The green box indicates our Red Clump Region. Dark remnant events with probability above 75 per cent are marked in red.} 
\label{fig:cmd}
\end{figure}

In order to limit our sample of events to those with sources at around 8 kpc (\eg \citealt{Pietrukowicz2015}), \ie in the Galactic Bulge, we constructed a colour-magnitude diagram (CMD), as shown in Fig. \ref{fig:cmd}. 
The $I$-band magnitudes of the sources were de-blended using $\Izero$ and $\fs$ from the microlensing parallax MCMC models and were corrected for extinction, based on \cite{Nataf} extinction maps for OGLE-III. 
We must note here, that due to the scarcity of $V$-band observations during the OGLE-III phase, the colour information was, in the case of most events, only available for the baseline. Hence our $V-I$ colour is only an approximate colour of the source. However, as already noted in \cite{Wyrzykowski2015}, the bulge Red Clump Giants (RCGs) are typically brighter than most of other stars on the CMD, hence even if blended with a Main Sequence star in the baseline, a RCG source would not change position greatly on the CMD.

We defined the Red Clump Region (RCR) as $0.5<(V-I)_0<2.5$ and $12<(I_\mathrm{source})_0<16$ for extinction-free magnitudes -- this selected 26 sources. 
For sources from the RCR we assumed their distance to be $\DS=8$ kpc.

The source in one of our events, PAR-42, has been observed spectroscopically with the 2.5m Ir\'{e}n\'{e}e du Pont telescope at Las Campanas Observatory in April 2014 (see \citealt{PietrukowiczSpec} for details). 
The low-resolution spectra, obtained with the B\&C spectrograph, clearly showed that the source is a K3 Red Clump Giant, confirming our conclusions based solely on the CMD from OGLE-III.

\subsection{Simulations and Detection Efficiency}

Our experiment is naturally not sensitive to every parallax event. For example we might have missed an anomaly due to parallax in an event during a gap in our observation season.
In order to correct for the detection efficiency we conducted simulations of parallax microlensing events.

The events were simulated in a similar fashion as described in \cite{Wyrzykowski2015}. 
Mock microlensing events were generated on top of light curves from the OGLE-III Bulge data base, preserving any intrinsic variability and original noise properties originating from varying observing conditions (seeing, airmass, background). Thanks to drawing random stars from the OGLE data base we reproduced the underlying observed luminosity function for the baseline of the simulated events. Blending parameters ($\fs$) were drawn from an empirical distribution obtained based on the comparison between observed OGLE stellar density and archival Hubble Space Telescope (HST) $I$-band images (as shown in \citealt{Wyrzykowski2015}).
The impact parameter ($\uzero$) was taken from a flat distribution from 0 to 1 and the time of minimum approach ($\t0$) was randomly chosen from the entire range of the covered epochs ($2100 < \mathrm{JD}-2450000 < 5000$). We simulated events with time-scales $\tE$ ranging from 1 to 1000 days, and parallaxes $\piE$ ranging from 0.0001 to 0.5, with a random angle for $\piE$, decomposed into North and East components ($\piEN$ and $\piEE$). 

After applying all cuts as for the real data (including Random Forest classifier), the efficiency was  derived as a function of two parameters: $\tE$ and $\piE$. 

\begin{figure}
\centering
\includegraphics[width=9cm]{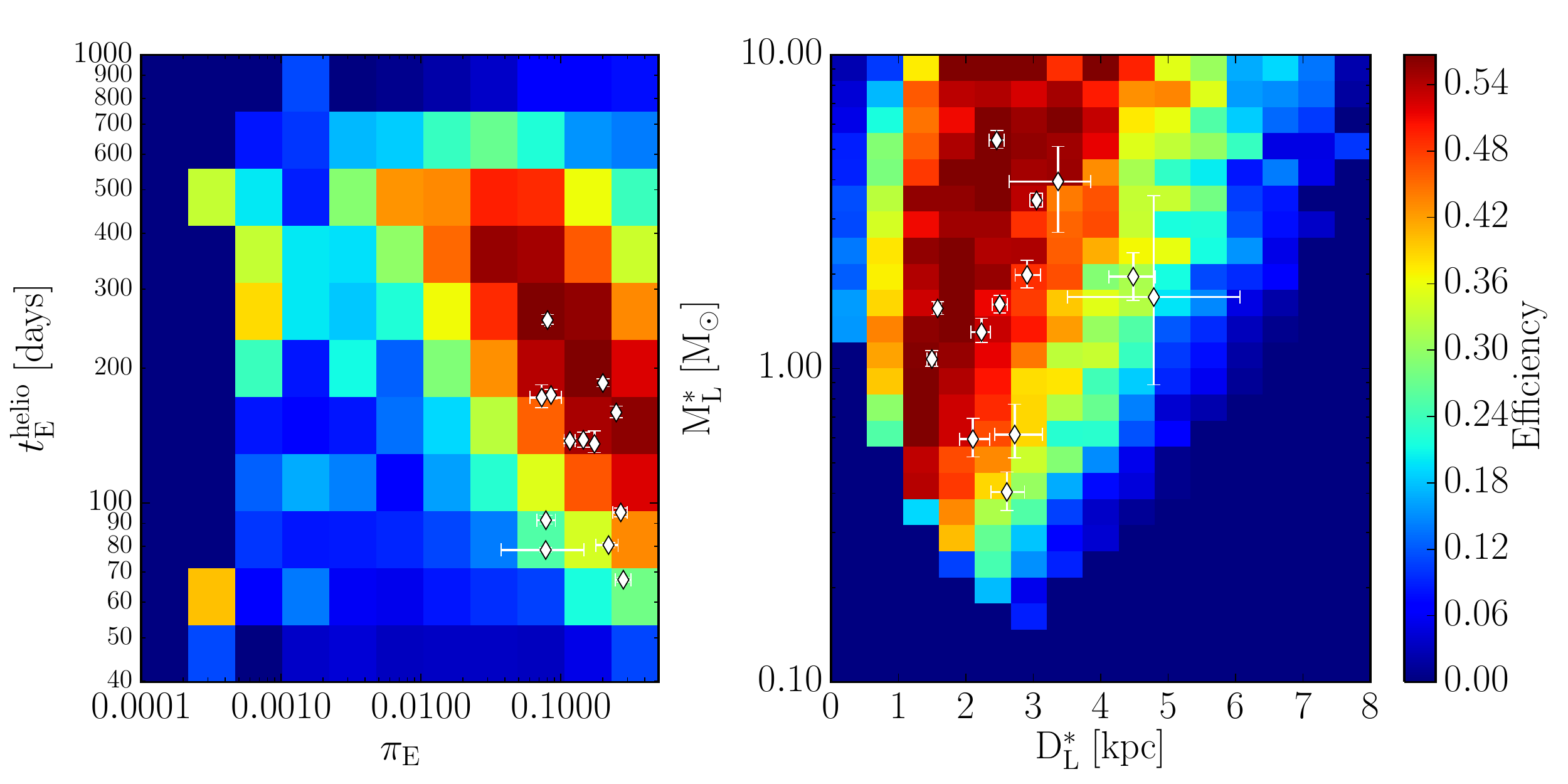} 
\caption{Detection efficiency for parallax events as a function of $\tEhelio$ and $\piE$ (left panel). Dots denote detected microlensing parallax events. 
Right panel shows the efficiency as a function of a mass and distance to the lens, for assumed, fixed proper motion of $\murel=5$ mas/yr. 
}
\label{fig:eff}
\end{figure}

Figure \ref{fig:eff} shows the derived detection efficiency as a function of $\tE$ and $\piE$. 
All parallax microlensing events detected in the OGLE-III data are marked as dots.
When combining both parameters, the efficiency can be also expressed as a function of plausible mass of the lens (right panel in Figure \ref{fig:eff}), when assuming a fixed value of $\murel=5$ mas/yr, the most likely value of relative proper motion (see next section).

As expected, we are not likely to detect the parallax signal in short time-scale events with small parallax modulation, hence, for low-mass lenses or lenses located in the Galactic bulge.
Decreasing efficiency for very long events and with very strong parallax is due to the fact that such events usually exceed the data span we covered in OGLE-III and their parallax modulations were so strong that there was hardly any baseline in the light curve to conclude on its microlensing nature. 
Our search for parallax events has the highest efficiency for long events or large parallaxes, reaching its maximum of about 55 per cent for time-scales of about 300 days and parallaxes $\sim0.1$. 
There is also an extension at very long time-scales ($\sim600$ days) for smaller $\piE\sim0.1$. 
In terms of approximate masses and distances, our pipeline is most sensitive for lenses heavier than about 0.7 $\msun$ and at a distance between 1 and 4 kpc. 
Our detection efficiency also remains high (more than 50 per cent) for very heavy lenses of about 10 $\msun$, where the simulation for efficiency stopped.

\section{Dark Remnant Candidates}
For a regular standard microlensing event it is typically impossible to obtain a unique solution for the distance and mass of the lens, since there is only one physical parameter ($\te$) obtained from the light curve model, which is linked with source and lens distance, lens mass and lens-source relative velocity. 
However, for parallax events the situation improves as we additionally measure the microlensing parallax. The mass and distance of a lensing object can be then derived from:

\begin{equation}
M=\frac{\theta_\mathrm{E}}{\kappa \pi_\mathrm{E}} = \frac{\mu_\mathrm{rel}^{\rm{helio}} \tEhelio}{\kappa \pi_\mathrm{E}}
\end{equation}

\begin{equation}
D_{\rm{L}}=\frac{1}{\mu_{\rm{rel}}^{\rm{helio}} \tEhelio \pi_{\rm{E}} + 1/D_{\rm{S}}}
\end{equation}

where we used the fact that the angular size of the Einstein radius can be rewritten as product of the length of the vector of the heliocentric relative proper motion $|\mu_\mathrm{rel}|=|\mu_L-\mu_S|$ 
between lens (L) and source (S) and the eventÕs time-scale $\te$. 
Proper motions and the time-scale should be measured in the same frame, either geo- or heliocentric. We therefore converted the time-scales of events obtained in our geocentric parallax model into the heliocentric frame, following \cite{Skowron2011} (in other words, the motion of the Earth was taken out from the relative proper motion computation, hence if the lens was observed from the Sun, it would have the time-scale of $\tEhelio$, hereafter denoted $\tE$).

\subsection{Proper motions}
Since we do not know the actual relative proper motions for our events, we had to assume it. 
In order to constrain the value of $\murel$, we assumed the sources belonged to the bulge population and had bulge-like distribution of velocities, and lenses are in the disk and follow disk's velocity and density distributions. 
We could then draw a random relative motion, composed of random source and lens motions, and then use the prior on how likely is that random relative proper motion given standard velocity and stellar density distributions for the bulge and disk (\eg \citealt{Batista2011}). 
The values were slightly adjusted to match the recently measured proper motions by \cite{Calamida2014} based on HST (see below).

Based on the CMD positions (Fig. \ref{fig:cmd}) of the sources we assumed that sources belong to the bulge population and are located at a fixed distance of 8 kpc. 
This assumption is also in agreement with the results of the simulation of NS and BH microlensing events done by \cite{Oslowski2008}.
The lenses, on the other hand, are distributed at various distances within the Galactic disk.
The closer the lens, the more likely a measurable parallax effect will be produced, hence we expect a bias towards nearby lenses.
Therefore, we assumed here bulge-like motions for the sources and disk-like motions for the lenses. 
Because the direction of the vector $\piEvec$ coincides with the direction of the relative proper motion $\murelvec$, we were able to narrow the selection of valid combinations of $\mu_S$ and $\mu_L$ using the angle defined by $\piEvec$. 

\subsection{Probability Density Functions}
MCMC parallax models for each event returned samples of solutions (sets of all microlensing parameters with parallax), whose densities reflect their probability. 
Mixing those values with lens and source proper motions, we obtained probability densities for masses and distances of our events, using Equations (2) and (3).
Note that any parallax model very often had two or more equally likely separate solutions (\eg \citealt{Gould2004}, \citealt{Smith2005}), therefore we carefully investigated MCMC results for each event to make sure the entire parameter space was explored and all possible solutions were discovered.

In order to transform the MCMC samples in microlensing parameter space to probability density functions (PDFs) for mass and distance, we transformed their density using the Jacobian and priors on the relevant parameters, following \cite{Batista2011}.
In our case, however, the relative proper motion ($\murel$) was one of the independent variables on the side of the microlensing parameters (as we provided its value in order to compute the mass). Each MCMC sample was weighted with the following weight on microlensing rate ($\Gamma$):

\begin{equation}
\frac{\rm{d}^4\rm{\Gamma}}{\rm{d}\tE\rm{d}\murel\rm{d}^2\piE}=
\frac{4}{AU}\nu(x,y,z)f(\murelvec)[g(M)M]\left(\frac{D_L\thetaE}{\tE}\right)^4\frac{\tE}{\piE}
\end{equation}

where $\nu$ is the stellar density distribution, $f(\murelvec)$ is the distribution of relative proper motions, $g(M)$ is the mass function. 
For the stellar distributions we assumed a thick Galactic disk with 0.6 kpc and 2.75 scale height and length, respectively. Velocities of the bulge sources in the galactic coordinates were assumed as (0,0) km/s in the heliocentric frame with a dispersion of (80,80) km/s. For the disk we assumed (0,200) km/s and a dispersion of (40,55) km/s. The mass function was assumed to be $\propto M^{-1.75}$.

\begin{figure}
\centering
\includegraphics[width=8cm]{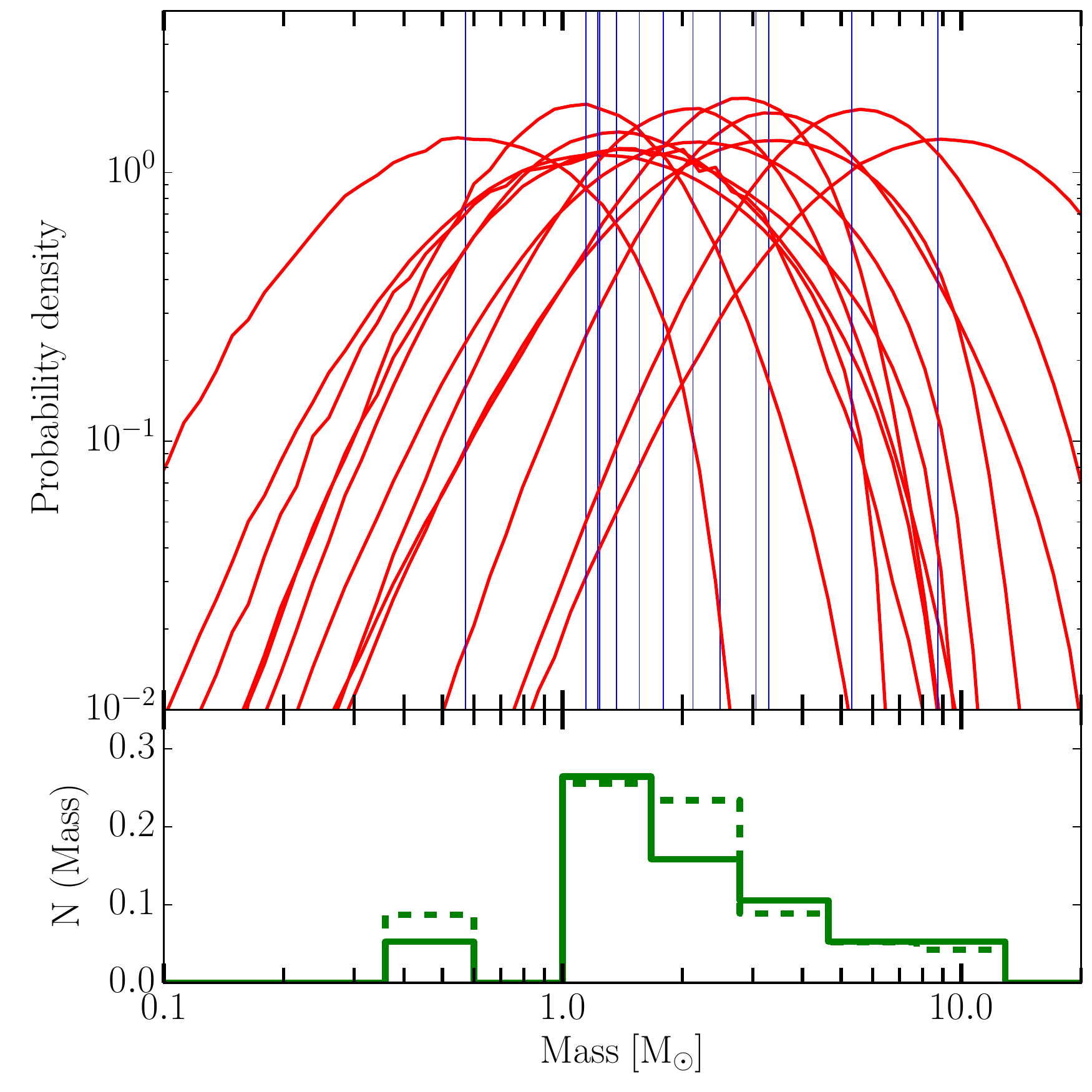} 
\caption{Probability density functions (PDF) for masses of 13 dark remnant candidates (red thick lines) as obtained for parallax microlensing events. Only the highest dark remnant probability solutions are shown. 
Thin vertical blue line for each PDF indicate its median value.
Bottom panel shows a normalised histogram of median masses as measured (solid line) and corrected for detection efficiency (dashed line).
}
\label{fig:pdfs}
\end{figure}

The individual PDFs for masses of all 13 dark remnant candidates are shown in Fig. \ref{fig:pdfs}. Only shown are mass distributions for the solution with the highest dark remnant probability. 
Lower panel in Fig. \ref{fig:pdfs} shows histogram of median masses of remnant candidates (solid line). 
The median mass distribution corrected for the detection efficiency is drawn with dashed line. 
Table \ref{tab:massesdistances} shows medians and 1 sigma errors (for masses and distances found for all Red Clump source lenses, $\ie$ the events for which those parameters were possible to estimate. 
The median values were computed from 1-D likelihood distributions of each parameter marginalized over all other dimensions.
The dark remnant candidates are included in that table.

\subsection{Dark Remnants Candidates}
Stellar remnants acting as lenses should be dark, \ie they should not contribute to the overall observed light. Microlensing allows us not only to constrain the mass of the lens, but also its light output.

Having drawn a relative proper motion for each sample in the MCMC solution, we computed the equivalent mass and distance of the lens, following Eqs. (2) and (3).
Then, for a given mass and distance, the luminosity of the lens was derived, using a luminosity-mass relation for main sequence stars: $L\sim M^4$ for masses less than 2 $\msun$ and $L\sim1.5\times M^{3.5}$ for $M>2\msun$.

\begin{figure}
\centering
\includegraphics[width=8cm]{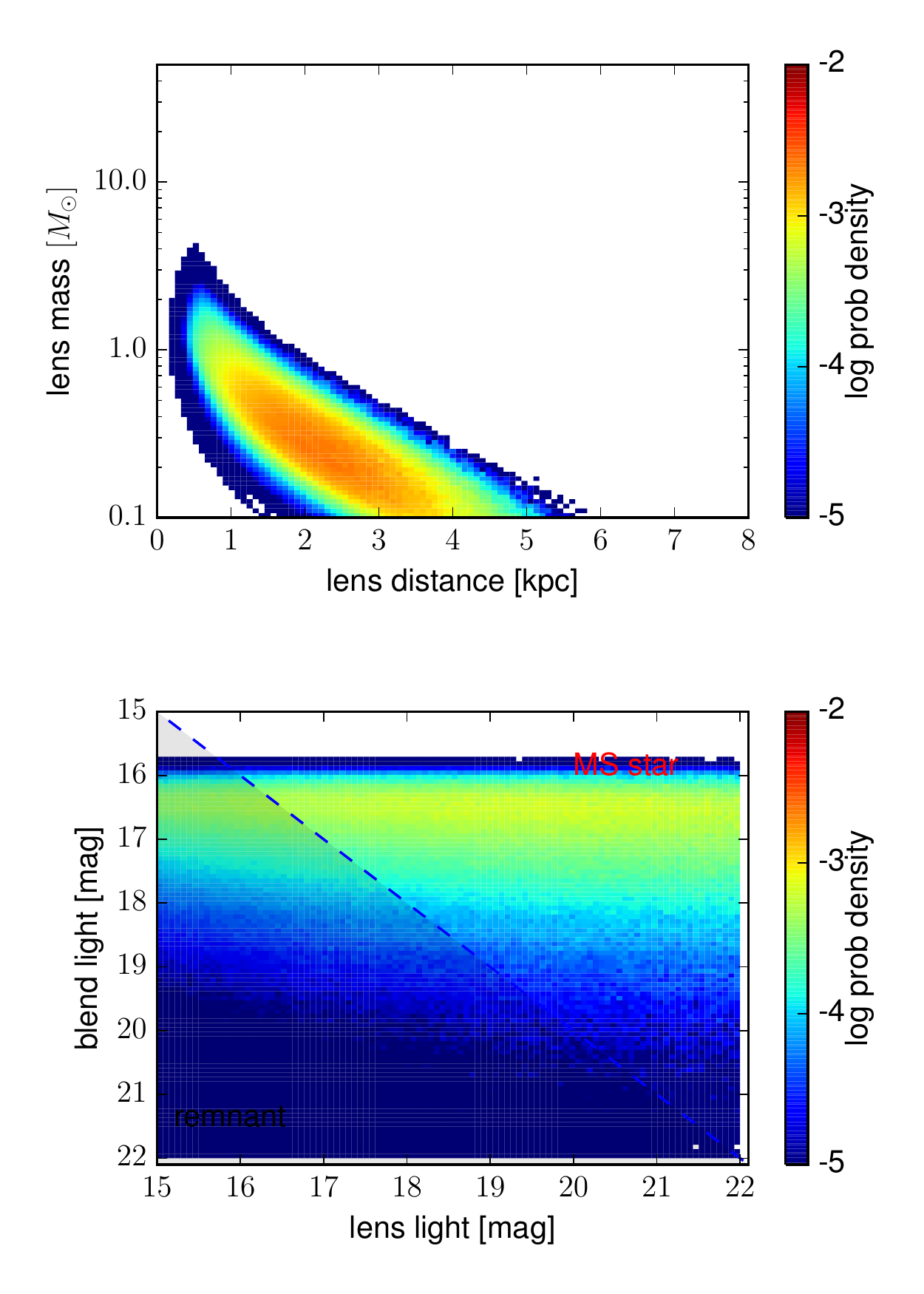} 
\caption{OGLE3-ULENS-PAR-42, an example of a non-remnant lens parallax event. Top: mass-distance probability density. Bottom: comparison of computed blended light and lens light expected for a given mass and distance. Shown is the solution with $\u0<0$ and probability 25.6 per cent for dark remnant (integral below dashed line). The lens is most likely a $\approx 0.4 \msun$ Main Sequence star at $\approx$ 2 kpc. 
}
\label{fig:massdist-par-42}
\end{figure}

The computed brightness of the lens was then compared with the amount of additional light (blended light) in the microlensing event, after subtracting the light of the lensed source, which is returned in the microlensing model as $\fs$. 
Note that in the crowded Bulge region, blending occurs even without microlensing, when the stars are closer to each other than a typical seeing radius ($\sim$1 arc sec in case of OGLE reference images), meaning that in case of microlensing events there will often be yet another (third) source of light, not related to the microlensing event. Hence, by attributing the entire blending light to the light of the lens, we place an upper limit on the flux originating from the lensing object. 
The brightness of the blend was corrected for the interstellar extinction using parameters from \cite{Nataf}. 
The bottom panel in Fig.~\ref{fig:massdist-par-42} shows a comparison between the expected brightness of the lens given its mass and distance and modelled blending light for a Main Sequence lens, OGLE3-ULENS-PAR-42.

We define the probability that the lens is a dark remnant as the ratio of the integrated density when there is not enough blending light to account for a luminous Main Sequence lens with a given mass  to the total density, \ie the integrated area in the bottom-left triangle in Fig. \ref{fig:massdist-par-42}.

Table \ref{tab:probs} lists all parallax events studied here and includes the probability for each of the events being a dark remnant for each solution. Note, some events had more than 2 degenerate solutions and probabilities were derived for each of them separately, as they yielded different blending parameters, masses and distances.


\begin{table}
\begin{scriptsize}
\centering
\caption{Red Clump source events and their probabilities of having a dark remnant lens. Probabilities are computed for all solutions and events are sorted according to the highest probability of any of the solutions. PAR-08 is the last one included in the 75 per cent threshold.}
\label{tab:probs}
\begin{tabular}{lllll}
\hline
Lens name  &  $u_0>0$ & $u_0>0$ & $u_0<0$ & $u_0<0$ \\
OGLE3-ULENS- &  $\piEN>0$ & $\piEN<0$ & $\piEN>0$ & $\piEN<0$ \\
\hline
\hline 
PAR-02 & 99.2 & - & - & 99.8\\
PAR-13 & - & 99 & 97.9 & -\\
PAR-05 & 95.4 & - & - & 94.1\\
PAR-07 & - & 92.1 & 85.9 & -\\
PAR-04 & 91.2 & - & - & 91.7\\
PAR-19 & - & 89.2 & - & 62.6\\
PAR-33 & 87.9 & 41.6 & 22 & 60.6\\
PAR-15 & - & 87.1 & 73.9 & -\\
PAR-09 & - & 86.9 & - & -\\
PAR-39 & - & 86.8 & - & 85.9\\
PAR-28 & 77.8 & - & - & 85.7\\
PAR-27 & 85.3 & 74 & 52.8 & 60\\
PAR-08 & 76.6 & - & - & 72.5\\
PAR-24 & 71.6 & 45.3 & 51.4 & 41.8\\
PAR-34 & 63.6 & - & 70.8 & -\\
PAR-03 & 65.2 & - & - & 61.5\\
PAR-22 & 57.2 & - & - & -\\
PAR-01 & - & - & 53.8 & -\\
PAR-44 & - & 53.2 & 50.3 & -\\
PAR-06 & 35.7 & - & - & 43.6\\
PAR-42 & - & 43.2 & 25.6 & -\\
PAR-36 & - & 42.5 & - & 32.1\\
PAR-16 & - & 33.4 & 40.9 & -\\
PAR-23 & - & 38.7 & 27.8 & -\\
PAR-47 & - & 33.3 & - & -\\
PAR-14 & 5.5 & - & - & 12.0\\
\hline
\hline
\end{tabular}
\end{scriptsize}
\end{table}

\begin{table}
\begin{scriptsize}
\centering
\caption{Most probable (1-D median and 1 sigma) masses, distances and total blend luminosities for all events with Red Clump sources. Multiple entries are shown for different solutions found, indicated in the last column. 
}
\label{tab:massesdistances} 
\begin{tabular}{lllll}
\hline
Lens name  &  mass & distance & blend & solution \\
OGLE3-ULENS- &  [$\msun$] & [kpc] & [mag] & ($\u0$ $\piEN$)\\
\hline
\hline
PAR-01 & $1.0^{+1.8}_{-0.6}$ & $1.3^{+1.1}_{-0.7}$ & $14.53^{+0.64}_{-0.03}$ & -  \\
PAR-02 & $8.7^{+8.1}_{-4.7}$ & $1.8^{+1.1}_{-0.8}$ & $16.16^{+0.76}_{-0.04}$ & -  \\
PAR-02 & $9.3^{+8.7}_{-4.3}$ & $2.4^{+1.1}_{-1.0}$ & $15.14^{+0.17}_{-0.04}$ & +  \\
PAR-03 & $0.9^{+1.3}_{-0.5}$ & $1.3^{+1.1}_{-0.8}$ & $16.82^{+0.46}_{-0.01}$ & -  \\
PAR-03 & $0.9^{+1.3}_{-0.5}$ & $1.4^{+1.1}_{-0.8}$ & $17.03^{+0.44}_{-0.01}$ & +  \\
PAR-04 & $1.6^{+1.7}_{-0.8}$ & $1.6^{+1.1}_{-0.8}$ & $15.50^{+0.71}_{-0.02}$ & -  \\
PAR-04 & $1.3^{+1.5}_{-0.6}$ & $1.7^{+1.2}_{-0.8}$ & $15.78^{+0.66}_{-0.01}$ & +  \\
PAR-05 & $4.8^{+4.0}_{-2.5}$ & $1.8^{+1.2}_{-0.7}$ & $15.37^{+1.25}_{-0.06}$ & -  \\
PAR-05 & $3.3^{+2.7}_{-1.5}$ & $2.9^{+1.1}_{-0.9}$ & $14.68^{+0.52}_{-0.02}$ & +  \\
PAR-06 & $1.0^{+1.3}_{-0.5}$ & $1.3^{+1.2}_{-0.7}$ & $14.39^{+0.17}_{-0.05}$ & -  \\
PAR-06 & $0.8^{+1.3}_{-0.5}$ & $1.6^{+1.2}_{-0.8}$ & $14.19^{+0.18}_{-0.06}$ & +  \\
PAR-07 & $3.0^{+2.2}_{-1.5}$ & $3.5^{+1.1}_{-1.0}$ & $15.62^{+0.05}_{-0.01}$ & -  \\
PAR-07 & $3.1^{+3.1}_{-1.6}$ & $2.1^{+1.3}_{-0.8}$ & $16.71^{+0.61}_{-0.02}$ & +  \\
PAR-08 & $1.2^{+1.3}_{-0.6}$ & $1.4^{+1.1}_{-0.7}$ & $15.48^{+0.79}_{-0.06}$ & -  \\
PAR-08 & $1.4^{+1.5}_{-0.7}$ & $2.0^{+1.1}_{-0.8}$ & $15.30^{+0.62}_{-0.04}$ & +  \\
PAR-09 & $1.2^{+1.4}_{-0.8}$ & $1.3^{+1.1}_{-0.7}$ & $16.63^{+0.65}_{-0.05}$ & +  \\
PAR-13 & $5.1^{+3.6}_{-2.5}$ & $2.8^{+1.1}_{-0.9}$ & $16.40^{+0.21}_{-0.03}$ & -  \\
PAR-13 & $5.3^{+3.9}_{-2.4}$ & $3.1^{+1.1}_{-1.0}$ & $16.33^{+0.31}_{-0.04}$ & +  \\
PAR-14 & $0.2^{+0.3}_{-0.1}$ & $1.1^{+1.1}_{-0.7}$ & $15.04^{+0.01}_{-0.01}$ & -  \\
PAR-14 & $0.1^{+0.1}_{-0.1}$ & $1.2^{+1.1}_{-0.8}$ & $15.38^{+0.02}_{-0.01}$ & +  \\
PAR-15 & $1.6^{+1.6}_{-0.8}$ & $2.2^{+1.2}_{-0.9}$ & $16.52^{+0.60}_{-0.04}$ & -  \\
PAR-15 & $2.1^{+2.3}_{-1.1}$ & $2.0^{+1.2}_{-0.8}$ & $16.55^{+0.63}_{-0.04}$ & +  \\
PAR-16 & $0.3^{+0.4}_{-0.2}$ & $1.7^{+1.1}_{-0.8}$ & $16.46^{+0.53}_{-0.02}$ & -  \\
PAR-16 & $0.5^{+0.6}_{-0.3}$ & $1.7^{+1.1}_{-0.8}$ & $16.11^{+0.82}_{-0.05}$ & +  \\
PAR-19 & $1.6^{+1.6}_{-0.7}$ & $3.1^{+1.3}_{-1.2}$ & $14.85^{+0.33}_{-0.01}$ & -  \\
PAR-19 & $1.8^{+1.5}_{-0.9}$ & $4.3^{+1.1}_{-1.2}$ & $14.80^{+0.23}_{-0.01}$ & +  \\
PAR-22 & $0.7^{+0.8}_{-0.4}$ & $1.8^{+1.1}_{-0.8}$ & $16.40^{+0.51}_{-0.04}$ & +  \\
PAR-23 & $0.3^{+0.4}_{-0.2}$ & $1.6^{+1.1}_{-0.8}$ & $15.77^{+0.38}_{-0.06}$ & -  \\
PAR-23 & $0.4^{+0.6}_{-0.3}$ & $1.4^{+1.2}_{-0.7}$ & $15.79^{+0.37}_{-0.05}$ & +  \\
PAR-24 & $0.7^{+0.8}_{-0.5}$ & $2.1^{+2.3}_{-1.0}$ & $15.04^{+0.31}_{-0.02}$ & - - \\
PAR-24 & $1.6^{+1.2}_{-0.8}$ & $4.1^{+1.0}_{-1.2}$ & $14.59^{+0.33}_{-0.05}$ & - + \\
PAR-24 & $1.4^{+1.1}_{-0.7}$ & $4.5^{+1.1}_{-2.4}$ & $14.77^{+0.27}_{-0.03}$ & + - \\
PAR-24 & $1.2^{+1.2}_{-0.6}$ & $3.6^{+1.1}_{-1.2}$ & $15.63^{+0.34}_{-0.01}$ & + + \\
PAR-27 & $0.8^{+0.8}_{-0.5}$ & $1.5^{+1.3}_{-0.7}$ & $17.05^{+0.25}_{-0.01}$ & - - \\
PAR-27 & $1.1^{+1.3}_{-0.8}$ & $2.5^{+1.5}_{-1.3}$ & $16.02^{+0.17}_{-0.06}$ & - + \\
PAR-27 & $1.8^{+1.3}_{-0.8}$ & $3.1^{+1.3}_{-1.4}$ & $16.27^{+0.12}_{-0.04}$ & + - \\
PAR-27 & $1.2^{+1.4}_{-0.6}$ & $2.8^{+1.2}_{-1.1}$ & $17.36^{+0.35}_{-0.01}$ & + + \\
PAR-28 & $0.6^{+0.5}_{-0.3}$ & $1.5^{+1.1}_{-0.7}$ & $17.87^{+0.65}_{-0.01}$ & -  \\
PAR-28 & $0.8^{+0.6}_{-0.4}$ & $2.5^{+1.1}_{-0.8}$ & $16.58^{+0.63}_{-0.04}$ & +  \\
PAR-33 & $0.8^{+0.7}_{-0.4}$ & $1.5^{+1.1}_{-0.7}$ & $17.56^{+0.42}_{-0.02}$ & - - \\
PAR-33 & $0.4^{+0.4}_{-0.2}$ & $1.9^{+1.2}_{-0.9}$ & $16.71^{+0.11}_{-0.01}$ & - + \\
PAR-33 & $0.6^{+0.7}_{-0.4}$ & $1.4^{+1.2}_{-0.6}$ & $16.20^{+0.15}_{-0.04}$ & + - \\
PAR-33 & $1.1^{+0.7}_{-0.6}$ & $3.5^{+1.0}_{-1.0}$ & $18.60^{+0.18}_{-0.00}$ & + + \\
PAR-34 & $1.3^{+1.1}_{-0.6}$ & $3.4^{+1.1}_{-1.1}$ & $14.37^{+0.31}_{-0.03}$ & -  \\
PAR-34 & $1.3^{+1.0}_{-0.7}$ & $3.3^{+1.1}_{-1.1}$ & $14.37^{+0.27}_{-0.03}$ & +  \\
PAR-36 & $0.7^{+0.6}_{-0.4}$ & $1.5^{+1.1}_{-0.6}$ & $15.49^{+0.02}_{-0.01}$ & -  \\
PAR-36 & $0.9^{+0.8}_{-0.5}$ & $1.7^{+1.2}_{-0.7}$ & $15.51^{+0.02}_{-0.01}$ & +  \\
PAR-39 & $2.2^{+1.5}_{-1.1}$ & $5.3^{+1.0}_{-1.4}$ & $16.76^{+0.02}_{-0.00}$ & -  \\
PAR-39 & $2.5^{+1.7}_{-1.2}$ & $5.4^{+1.0}_{-1.4}$ & $16.77^{+0.02}_{-0.00}$ & +  \\
PAR-42 & $0.4^{+0.4}_{-0.2}$ & $1.8^{+1.1}_{-0.8}$ & $16.26^{+0.55}_{-0.05}$ & -  \\
PAR-42 & $0.5^{+0.4}_{-0.3}$ & $2.1^{+1.1}_{-0.8}$ & $16.34^{+0.44}_{-0.05}$ & +  \\
PAR-44 & $0.7^{+0.6}_{-0.4}$ & $2.8^{+1.1}_{-1.0}$ & $15.25^{+0.28}_{-0.05}$ & -  \\
PAR-44 & $1.0^{+0.9}_{-0.5}$ & $3.0^{+1.3}_{-1.1}$ & $15.22^{+0.34}_{-0.06}$ & +  \\
PAR-47 & $0.4^{+0.4}_{-0.2}$ & $2.2^{+1.1}_{-0.9}$ & $15.34^{+0.43}_{-0.07}$ & +  \\
\hline
\hline
\end{tabular}
\end{scriptsize}
\end{table}

\section{Discussion}
There has been more than 15,000 microlensing events detected so far since the beginning of the microlensing surveys in early 1990s.
\cite{Wyrzykowski2015} compiled a list of 3,500 pure standard microlensing events from OGLE-III data (2001-2009), the largest homogeneous sample of standard events. Those events prove useful for statistical studies of mass distribution in the Galaxy, however, they are not suitable for discovering individual remnant lenses since they provide only one parameter used in mass determination ($\te$).
Here we have carefully selected those long events which exhibited parallax signatures due to Earth motion, and were free of any other additional effects, (\eg binary lens), as robust measurements of $\piE$ guarantees a good mass estimate. 

By studying the light curves of 150 million sources monitored frequently for 8 years by the OGLE-III project we have found 13 candidates for dark stellar remnants acting as lenses in microlensing events, with probabilities higher than 75 per cent. There would be 19 candidates given a 50 per cent probability threshold and 5 for a conservative 90 per cent threshold.

For each candidate the mass and distance of the lens plus its light, to exclude regular stellar lenses, were estimated based on the microlensing parallax model fit to the light curve (for parameters $\te$ and $\piE$), and the unknown relative proper motion was drawn from observed empirical distributions. We therefore obtained probability density functions for the mass and distance of each remnant, as well as an accumulated probability that a given lens is non-luminous.

\subsection{Proper Motions and Distances}
The relative proper motion (PM, or $\murel$) between the source and the lens is the key missing ingredient for an exact determination of the mass and distance of the lens. 
Therefore, we have used the prior on distribution of PM measured for Galactic bulge and disk stars \citep{Calamida2013}, assuming our sources are in the bulge and the lenses in the disk.

However, if a source and lens were both located in the Galactic disk, they might together have very small value of $\murel$ if their projected motion is almost parallel. Such PM would decrease the measured mass of the lens (Eq. (2)) and would place the lens at a further distance (Eq. (3)), hence increasing a chance the lens is a normal disk star. 

Since the microlensing phenomenon amplifies the light of a distant source star, it is possible to determine the true brightness of the source, despite severe blending conditions typical for the bulge region (see \eg \citealt{WozniakPaczynski1997}, \citealt{Wyrzykowski2006}, \citealt{Smith2007}). 
Our sample of parallax events (59) was limited to only those with a source located in the Red Clump Region (26 events). 
Since the vast majority of Red Clump Giants towards the bulge are located at about 8 kpc, we assumed that our RCR sources have $\DS=8$ kpc and they follow the PM distribution of the Bulge stars. There are of course rare cases of RC stars in the Galactic disk, but they would be shifted on the CMD as brighter and less affected by extinction, hence might fall out of our RCR box. The same applies (but with the inverse argument about CMD location) to even rarer examples of Red Clump stars from the far side of the Galactic bulge. 
Note that our mass determination (Eq. (2)) does not depend on the assumption on $\DS$, but only on $\murel$. The distance to the lens, computed from Eq. (3), for a typical time-scale in our sample of $\sim$200 days and parallax $\piE\sim0.05$, can vary by at most
 0.5 kpc if $\DS$ is different from 8 kpc by $\pm$1 kpc, for small values of $\murel\sim2$. The error-bar from the MCMC modelling for $\DL$ is similarly wide. 

Another crucial assumption is that the lenses are located in the disk and follow the PMs of the disk stars. 
However, the events we use they exhibit a clear microlensing parallax signal and a large parallax signal  indicates a nearby lens. For our assumed $\DS=8$ kpc and typical $\piE\sim0.05$ most lenses should be located at $\DL<6$ kpc, hence clearly lying in the Galactic disk. 
On the other hand, a lens located in the Bulge would have a small and hard to detect $\piE$.
A relative PM distribution for such microlensing event would not differ much from the bulge-disk one, but even an increased chance for a smaller $\murel$ would be compensated by small value of $\piE$ and the mass measurement would not be affected significantly. 

Of course, our remnant lens candidates, especially black holes, do not have to follow general disk motions at all - for example they might reside in the Galactic halo or be moving as a result of a significant birth kick \citep{Fryer2012}. However, in such situations, their relative motion with respect to the source star would be even larger than assumed for the disk. Hence, our disk motion assumption provides a safe lower limit on the mass of the remnant. 

We have to note that the PDFs for the mass measurements are still wide and range often over an order of magnitude. 
If the assumptions on relative proper motions are invalid in case of some events, their masses could differ significantly. 
In principle, the lower mass end is heavily populated by white dwarfs, the distribution of which peaks at around 0.6 $\msun$ \citep{Kepler2007WD}. 
Most white dwarf lenses with absolute magnitude in range between 2-15 mag could still fit within the light of the blend of most of our candidates, at least in the tails of the probability distributions.
We note that our mass distribution does not show any prominent peak at masses where a vast majority of white dwarfs should be. Our experiment is less sensitive for lower masses (see Fig. \ref{fig:eff}) and the detection efficiency of our search pipeline (which does not include expected Galactic remnants population densities) reflects that (see dashed line in Fig. \ref{fig:pdfs}, however, there still should be significantly more WDs than neutron stars and black holes. This indicates that there still could be a strong bias in parallax microlensing events for heavier lenses and our search is still not completely sensitive to white dwarf lenses.

\subsection{Individual events}
Below we discuss some of the individual microlensing events having a candidate dark remnant lens.

\subsubsection*{OGLE3-ULENS-PAR-01} 
\begin{figure}
\centering
\includegraphics[width=8cm]{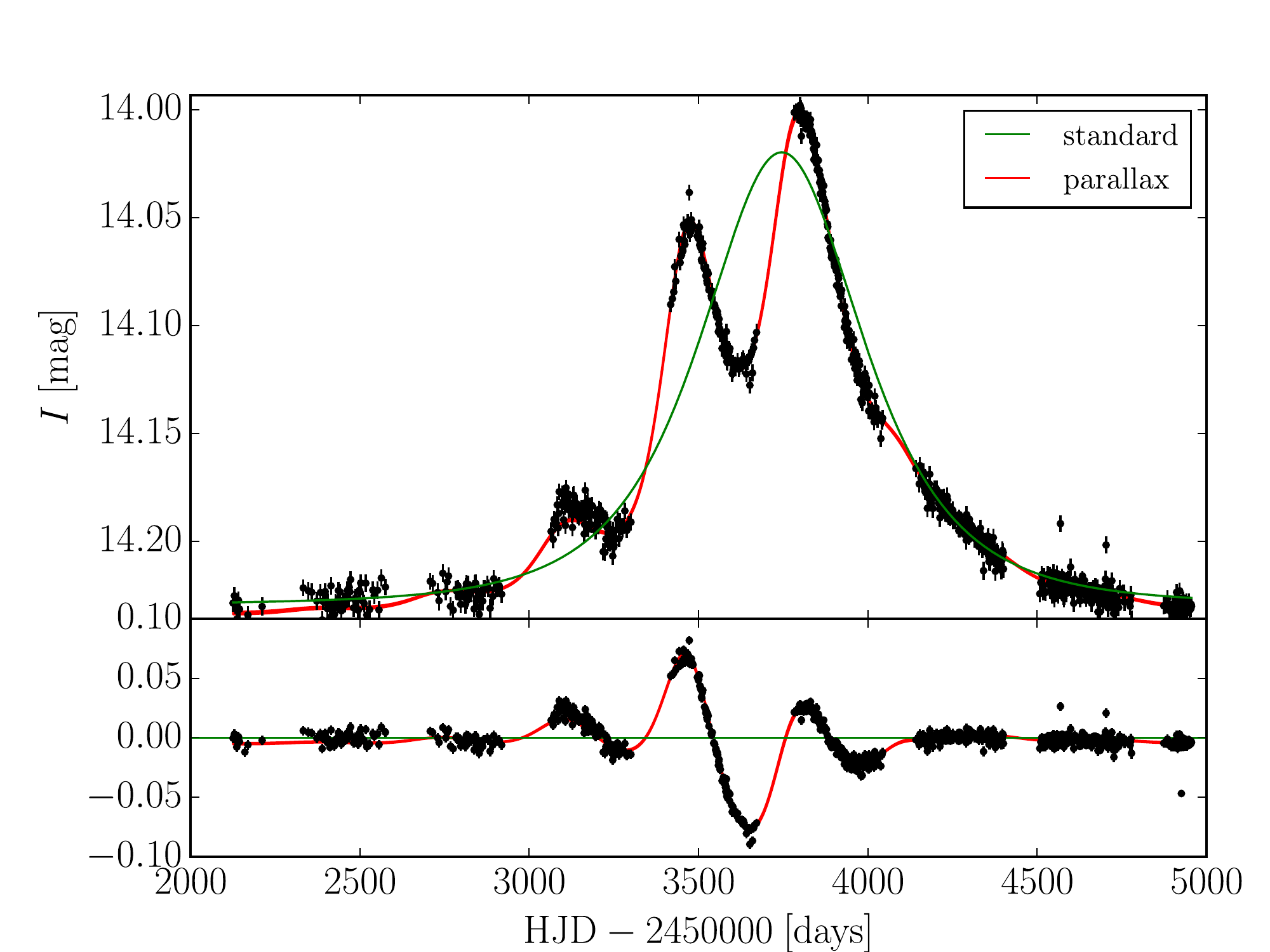}
\caption{Light curve and standard (green) and parallax (red) microlensing models and their residuals for the longest lasting multi-peak event OGLE3-ULENS-PAR-01. There was only one solution found and it has an equivocal 53.8 per cent probability for a 1.0 $\msun$ dark lens at about 1.3 kpc. The width of the parallax model curve (red) indicates 1 $\sigma$ uncertainty in the model.
}
\label{fig:vistulalc}
\end{figure}

OGLE3-ULENS-PAR-01 is the event with the longest duration and is probably the longest event ever observed lasting 2840.8 days ($\sim$7.8 years). The event was not found by the real-time OGLE-III data analysis (EWS), probably because its deviation had started already in 2002. In its light curve (shown in Fig. \ref{fig:vistulalc}) it exhibits multiple peaks due to the parallax effect modulation, making PAR-01 the second multi-peak microlensing event after OGLE-1999-BUL-19 \citep{Smith2002}. The large amplitude parallax signal immediately indicates that the lens is nearby. 
There was only one physically correct solution for parallax ($\u0<0$) returned by the MCMC modelling. 
MCMC modelling corrected for Galactic priors returned the most likely distance of just 1.3 kpc and mass of 1.0 $\msun$. 
At such a mass and distance and with a blending parameter of $\fs=0.68$ there was somewhat a low probability (53.8 per cent) that the lens is non-luminous.
At this moment we can not rule out between a main sequence star or a neutron star, however, in the tail of the probability distribution there is still a possibility of a massive white dwarf, if the relative proper motion between the bulge source and the nearby lens was significantly smaller than few mas/yr.  
On the other hand, since the lens is near (consistent with the large parallax signal), its proper motion could be different than that assumed for standard disk stars. 
If significantly higher, then the mass of the lens would also become significantly heavier, hence it would be more difficult to explain the blended light with a nearby massive main sequence star and a nearby neutron star and even a black hole solution could be considered.

Further observations of this object, including high resolution imaging, could potentially help resolve the source and the lens and a non-detection of a lens could indicate another dark remnant. 

\begin{table}
\begin{scriptsize}
\centering
\caption{Parallax model parameters of OGLE3-ULENS-PAR-01, the longest parallax microlensing event with multiple peaks in its light curve.}
\label{tab:paramsvistula}
\begin{tabular}{lll}
\hline
$\tzeropar$ & 3799.8 & [days]\\
 & & \\
$\t0$ & $3792.4^{+2.6}_{-2.4}$ & [days]\\ 
 & & \\
$\tEgeo$ & $264.4^{+15.0}_{-16.7}$ & [days]\\
 & & \\
$u_{\mathrm{0}}$ & $-0.98659^{+0.08509}_{-0.10841}$ & \\
 & & \\
$\piEE$ & $0.132^{+0.006}_{-0.005}$ & \\
 & & \\
$\piEN$ & $0.218^{+0.011}_{-0.009}$ & \\
 & & \\
$I_{\rm{0}}$ & $14.235^{+0.000}_{-0.000}$ & [mag]\\
 & & \\
$\fs$ & $0.68^{+0.17}_{-0.11}$ &\\
 & & \\
$\tEhelio$ & $324.7^{+19.0}_{-21.0}$ & [days]\\
 & & \\
$\chi^2/ndof$ & 2.06 & \\
 & & \\
 ndof & 874 &\\
\hline
\end{tabular}
\end{scriptsize}
\end{table}

\subsubsection*{OGLE3-ULENS-PAR-02} 
\begin{figure}
\centering
\includegraphics[width=8cm]{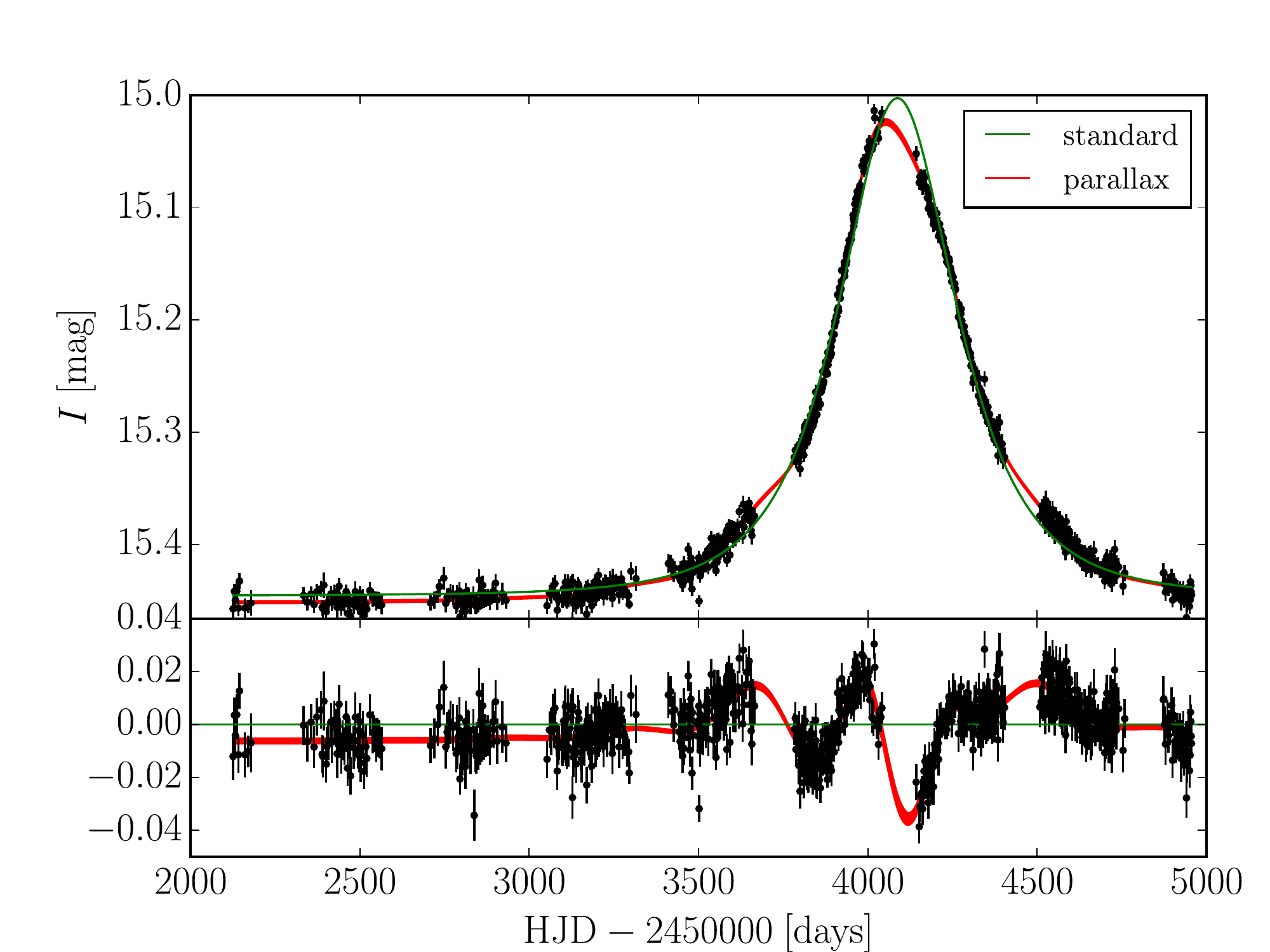}
\caption{Light curve and standard (green) and parallax (red) microlensing models and their residuals for the event OGLE3-ULENS-PAR-02. The solution shown ($\u0<0$) has a 99.8 per cent probability of a dark lens of 8.7$\msun$ lens at 1.8 kpc.
}
\label{fig:odralc}
\end{figure}

OGLE3-ULENS-PAR-02 is the second longest event in our sample, with a length of 2306.4 days (6.3 years), spanning until the very end of the OGLE-III phase in 2009. A small degree of amplification is also still visible in the OGLE-IV data from 2010. The light curve with the most probable microlensing model of PAR-02 is shown in Fig. \ref{fig:odralc}. 

\begin{figure}
\centering
\includegraphics[width=8cm]{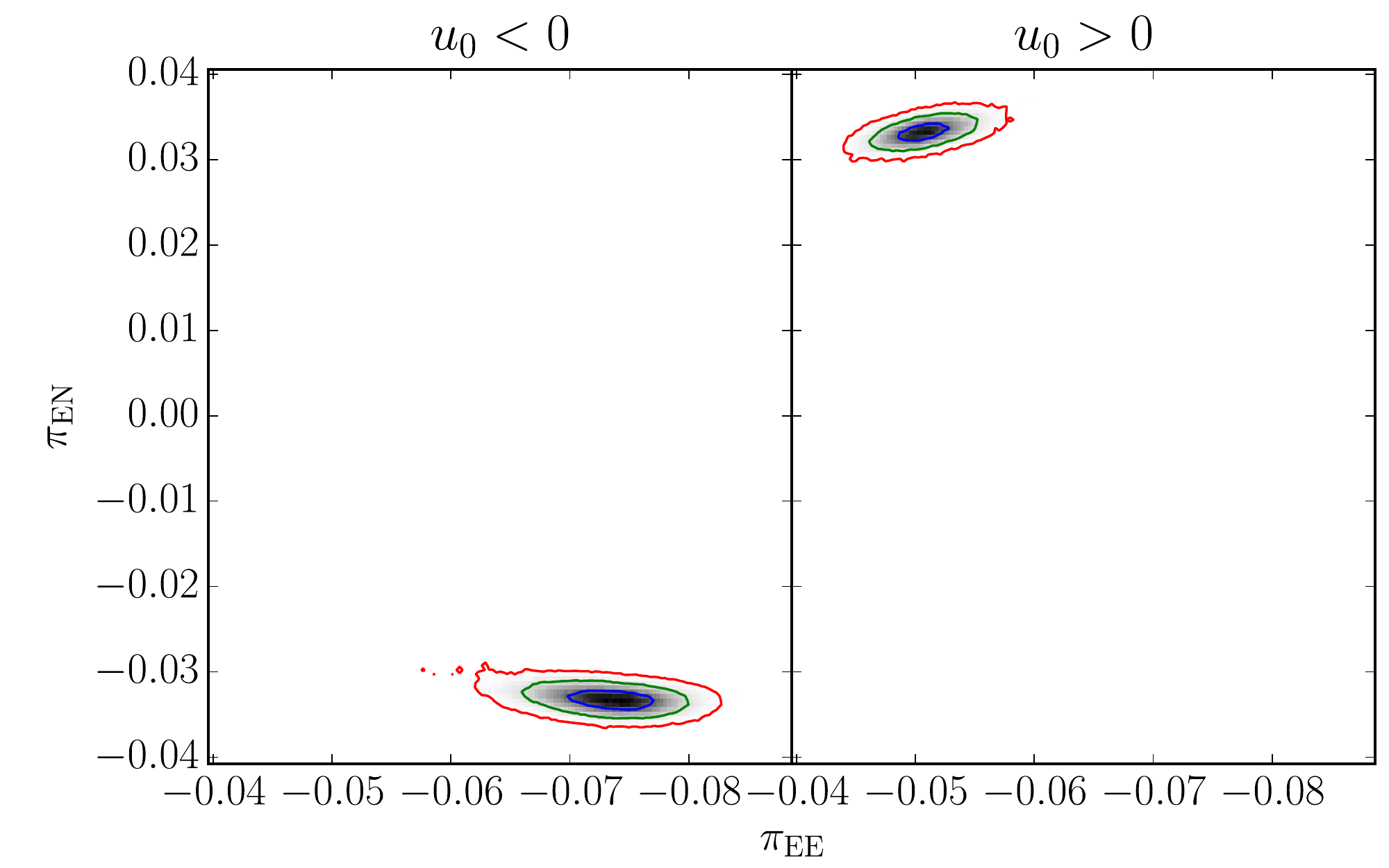}
\caption{Parallax degeneracy found in the photometric data parallax models for OGLE3-ULENS-PAR-02 event. Blue, green and red contours indicate 1,2,3 sigmas, respectively.}
\label{fig:odra-pi}
\end{figure}

The MCMC modelling yielded two degenerate solutions for $\uzero>0$ and $\u0<0$ (see Table \ref{tab:mcmc-rem}), with heliocentric time-scales of $\tE=296^{+8}_{-7}$ days and $\tE=256^{+7}_{-5}$ days. Parallax parameter $\piE$ solutions are shown in Fig. \ref{fig:odra-pi}. 

The light curve for this event is shown in Fig. \ref{fig:odralc} along with the standard and parallax microlensing models. The deviation from the standard model, shown at the bottom, has an amplitude of only $\pm$3 per cent, however, thanks to frequent monitoring and good photometry, the parallax anomaly is clearly visible with yearly fluctuations and is well reproduced by the model. 

\begin{figure}
\centering
\includegraphics[width=8cm]{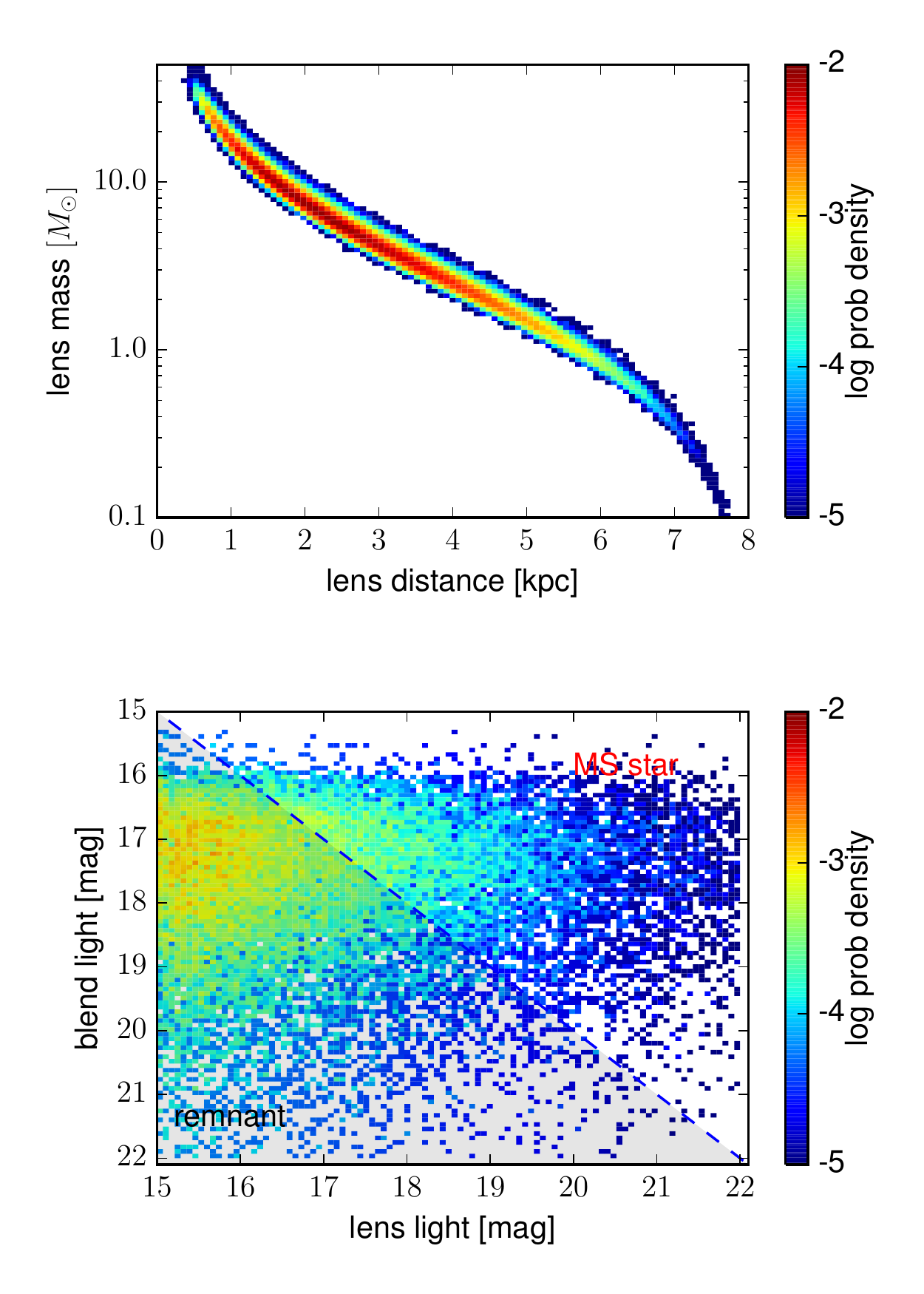}
\caption{OGLE3-ULENS-PAR-02, the most massive and most probable isolated black-hole candidate. Top: mass-distance probability density. Bottom: comparison of computed blended light and lens light expected for a given mass and distance. Shown is solution with $\u0<0$ and probability 99.8 per cent for dark remnant (integral below dashed line). The lens light axis stretches to the very bright magnitudes. 
}
\label{fig:massdist-odra}
\end{figure}

The two degenerate solutions differed significantly in the computed relative source light contribution to the overall light: $\fs=1.01^{+0.07}_{-0.08}$ and $\fs=0.64\pm0.04$, with a slight preference for the $\fs=1.01$ solution (that preference corresponding to a difference in $\chi^2$ of $\Delta\chi^2=8.5$).
A blending parameter close to 1.0 means that the source dominates the light at the place of the event and indicates very little room for any extra light from the blend, in particular from the lens. 
The distribution of allowed blend brightness in our solution is shown in the lower panel of Fig. \ref{fig:massdist-odra}. 
The blend can still have a brightness of about 17.7 mag, but for our most likely mass and distance, this most likely is not the lens itself. 

OGLE3-ULENS-PAR-02 has the highest probability for being a dark remnant among our candidates - both parallax solutions indicate probabilities above 99 per cent that the lens is a dark remnant.
The somewhat more probable of the two solutions ($\u0<0$) gives the lens at a distance of 1.8 kpc and with a mass of
$8.7^{+8.1}_{-4.7} \msun$, but the other solution (with some blending) yielded an even higher mass of $9.3^{+8.7}_{-4.3} \msun$ at 2.4 kpc. 
PAR-02 is the heaviest of our dark remnant candidates, hence it is most likely an isolated black hole. 
Interestingly, this black hole candidate is the only one in our sample where the lens mass is above the mass cut-off observed in black holes in X-ray binaries \citep{Ozel2010}. 
On the other hand, the detection efficiency in our pipeline for microlensing events with parameters similar to PAR-02 is relatively low (see Fig. \ref{fig:eff}), since the smaller value of $\piE$, the more massive the lens. 
Hence there might be more black hole lenses with masses above 6 $\msun$, which do not exhibit strong parallax signals in our data and are hidden among the standard microlensing events. 
Fig. \ref{fig:massdist-odra} shows the probability density for mass and distance for this solution as well as a  comparison between computed blend light with expected luminosity of the lens for its mass and distance. 

The mass-distance-blend light derived for the other solution ($\u0>0$ and$\fs=0.64\pm0.04$) also indicates a dark remnant/black hole with very high probability (99.2 per cent). There is blended light in this solution, however, it is most likely to be coming from an unrelated nearby blending star since a main sequence star with mass of 9.3 $\msun$ at distance of 2.4 kpc should be significantly brighter. The total blend light is estimated at $I_\mathrm{blend} \approx 16.3$ mag (extinction corrected). 

If, nevertheless, the lens was indeed the blend in this solution, given its brightness and assuming again a Main Sequence star from the disk, the lens would need to have a mass in the range between 0.9 and 1.1 $\msun$ and then the relative proper motion would need to be about 0.9 mas/yr. Such a small value of proper motion would mean that both source and lens move almost in parallel (projected). This is not impossible, but the chance is extremely small for bulge and disk stars. Such an interpretation can be tested with high resolution Adaptive Optics (AO) observations of the system after 10 or more years, when the source and lens would have separated by $\approx$10 mas (\eg \citealt{KozlowskiHST}, \citealt{Pietrukowicz2012M22}, \citealt{Batista2015}). With both source and lens being relatively bright, a deformation in the point-spread function should allow us to measure such delicate blend. On the other hand, a non-detection of lens light would allow us to rule out this solution with high confidence.

Since both solutions indicate a massive lens, the event OGLE3-ULENS-PAR-02 is the best black hole candidate in our sample, with its mass of 8.7 or 9.3$\msun$, depending on solution, and strong constraints on the lens light.
Nevertheless, further detailed AO and X-ray follow-up is necessary in order to rule out other solutions and to potentially detect the BHÕs interaction with the interstellar medium (\eg \citealt{AgolKamionkowski2002}, \citealt{Fender2013}). 

\subsubsection*{OGLE3-ULENS-PAR-05} 
\begin{figure}
\centering
\includegraphics[width=8cm]{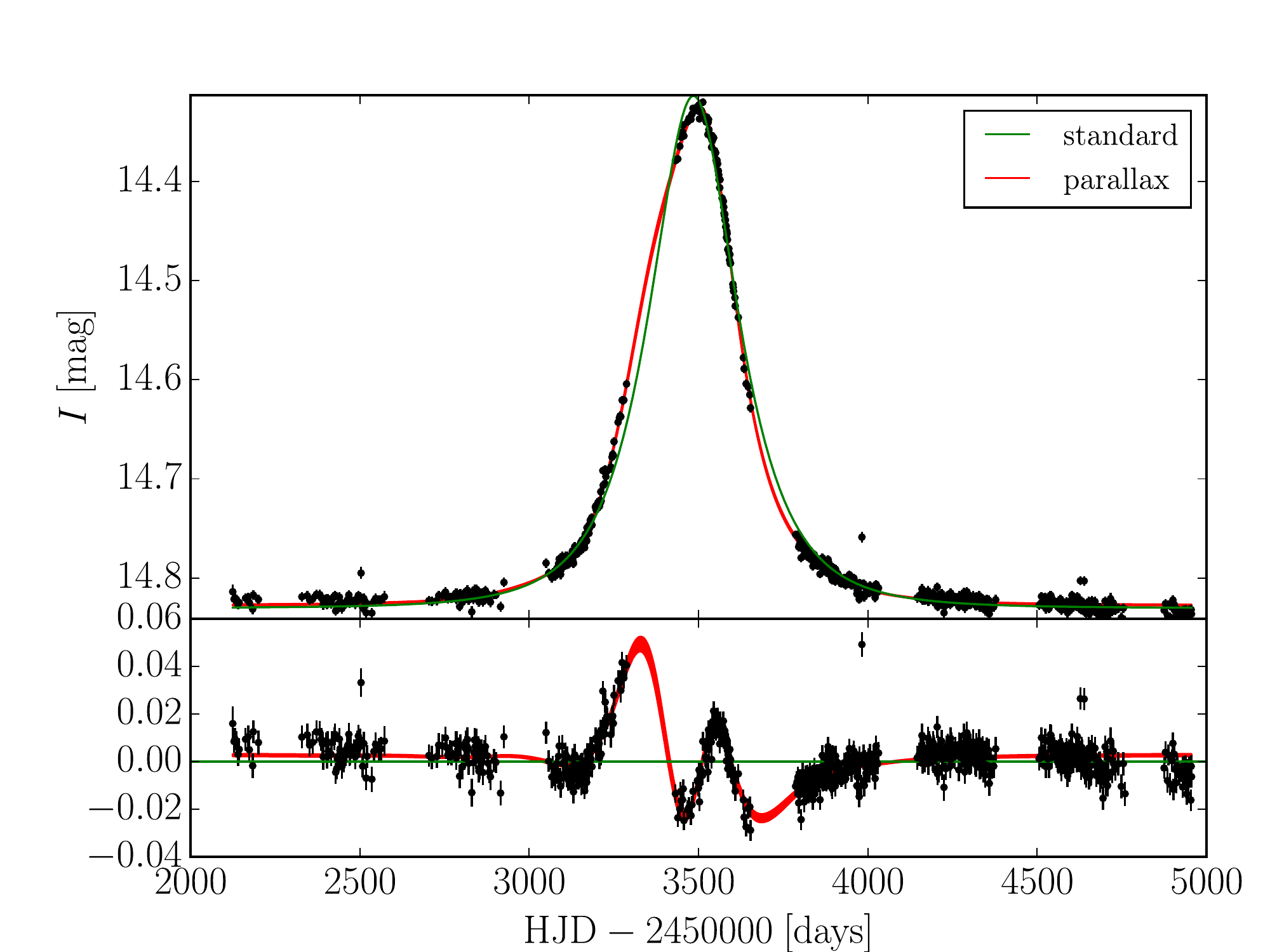}
\caption{Light curve and standard (green) and parallax (red) microlensing models and their residuals for the event OGLE3-ULENS-PAR-05. The solution shown ($\u0>0$) has a 95.4 per cent probability of a 3.3 $\msun$ lens at 2.9 kpc.
}
\label{fig:narewlc}
\end{figure}

\begin{figure}
\centering
\includegraphics[width=8cm]{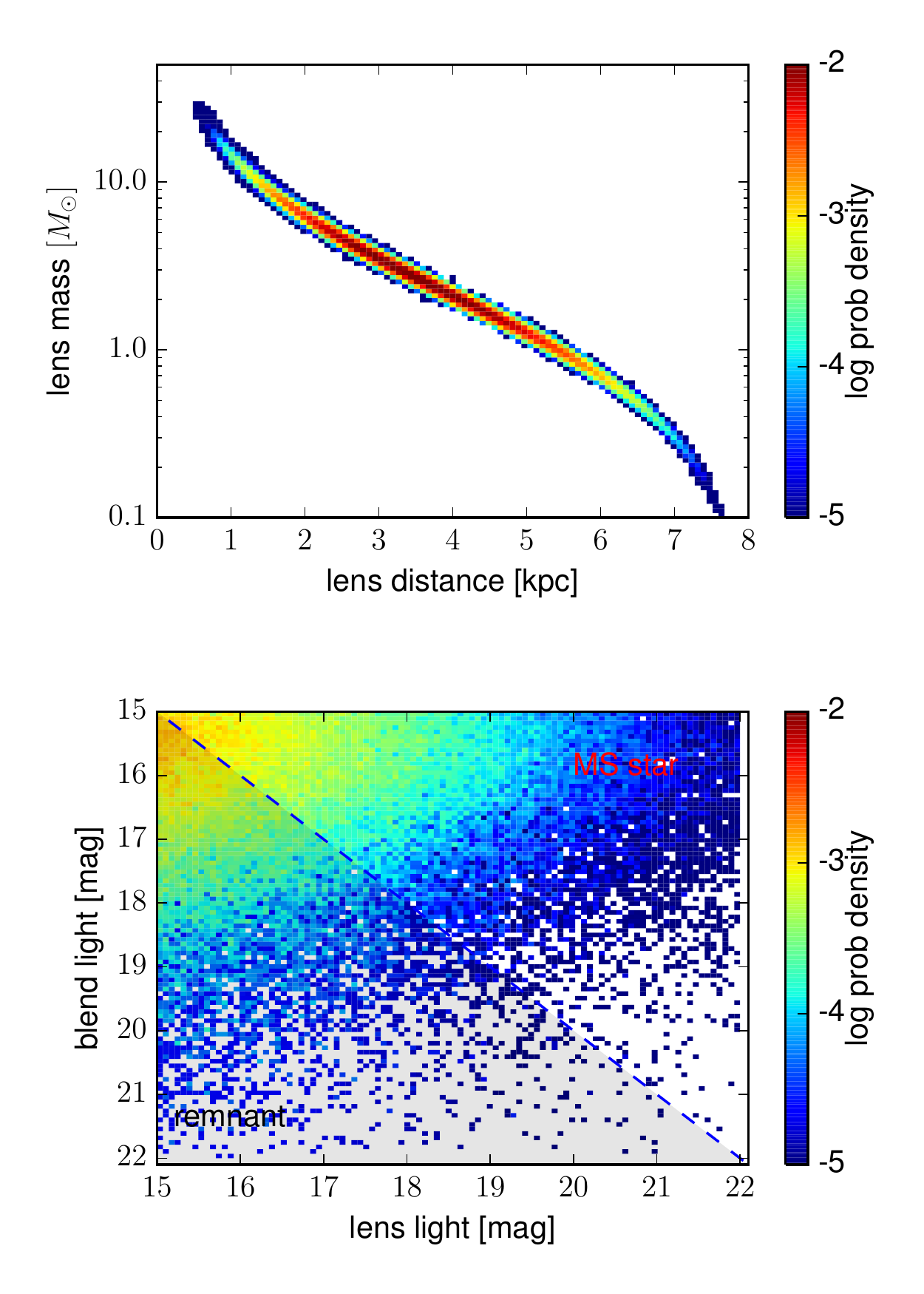}
\caption{OGLE3-ULENS-PAR-05, a dark remnant lens candidate from the mass gap. Top: mass-distance probability density. Bottom: comparison of computed blended light and lens light expected for a given mass and distance. Shown is solution with $\u0>0$ and probability 95.4 per cent for dark remnant (integral below dashed line). 
}
\label{fig:massdist-narew}
\end{figure}

OGLE3-ULENS-PAR-05 is another very strong candidate for a dark remnant lens. Its lightcurve, shown in Fig. \ref{fig:narewlc}, shows a very long-term and strong parallax anomaly, allowing for a good measurement of $\piE$. 
There were two solutions found. Both solutions have blending $\fs$ very close to 1, indicating no blending and leaving no room for light from a massive lens. Probabilities for dark remnant lens were above 95 per cent for both solutions. 
The most probable mass of the lens was estimated at $3.3^{+2.7}_{-1.5}$ and $4.8^{+4.0}_{-2.5} \msun$, for positive and negative $\u0$ solutions, respectively. 
Figure \ref{fig:massdist-narew} shows the mass-distance probability density and comparison of computed blended light and lens light.

If the mass of this dark remnant was indeed as our most likely estimate, this object would be residing in the very gap between known neutron stars and black holes. 
\cite{Kiziltan2013} suggested that the mass distribution of NSs, despite its apparent concentration between 1 and 1.5 $\msun$, can also spread to higher masses, also above 2 $\msun$. 
On the other hand, the mass distribution for black holes, found in X-ray binaries, have a cut off at about 6 $\msun$ \citep{Ozel2010}. 
Microlensing parameters of OGLE3-ULENS-PAR-05 place it in the region of relatively low detection efficiency (see Fig.~\ref{fig:eff}). 
A detection of OGLE3-ULENS-PAR-05 indicates there might exist a population of isolated NSs or BHs present in the mass distribution gap, in agreement with current theoretical predictions \citep{FryerKalogera2001}. 

\subsection{Mass distribution} 
Fig. \ref{fig:pdfs} shows individual mass probability density functions for the most probable remnant solution of our 13 events with remnant lens candidate. 
Median masses, listed also in Table \ref{tab:massesdistances}, are marked with blue vertical line and their distribution is shown in the lower panel of Fig. \ref{fig:pdfs}. 
It can be seen that correcting for the detection efficiency only mildly modifies the shape of the histogram, with the lowest and the highest masses being detected with somewhat lower detection efficiency. 
If indeed all our 13 lenses are due to remnants and we use their most probable masses, the derived approximate mass distribution has a peak between 1 and 2 $\msun$, most likely due to a mixture of single neutron stars and massive white dwarfs. 
The low-mass end is then populated by white dwarfs, however, as noted above, we seem to be not complete for WDs.
In the high-mass end of the distribution we see a continuum of masses, from 1 to 10 $\msun$. 
Strikingly, the shape of this mass function resembles the synthetic one obtained by \cite{FryerKalogera2001}, with a steady decline of masses of neutron stars and black holes, with no mass gap between the two populations of compact remnants. 

\subsection{Astrometric microlensing}
Microlensing events with heavy and relatively nearby lenses will have large angular Einstein radii, $\thetaE$, where $\thetaE \propto  (M \pirel)^{1/2}$ and $\pirel = AU(1/\DL - 1/\DS)$.
As shown in, \eg \cite{Walker1995}, \cite{Hog1995} and \cite{Dominik2000}, 
the motion of unresolved images in a microlensing event produces a characteristic displacement of the observed centroid of light of the source star of an order of $\thetaE$. 
An astrometric deviation from the unperturbed motion of the source is computed as 
$\delta(\vec{u}) = \frac{\vec{u}}{u^2+2}\thetaE$, where $\vec{u}$ is the vector between source and lens positions. 
It reaches its maximum amplitude of $\delta_\mathrm{max} \approx 0.7~ \thetaE$ for $\u0=\sqrt2$. 
It means that by knowing $\vec{u}$ from photometry and measuring the centroid displacement one can immediately derive the value of $\thetaE$ for microlensing events without the need of any other additional effects (finite source or caustic crossing). 
Following Eq. (1), having measured both $\piE$ and $\thetaE$ one can compute the mass of the lens directly. 

\begin{figure}
\centering
\includegraphics[width=8cm]{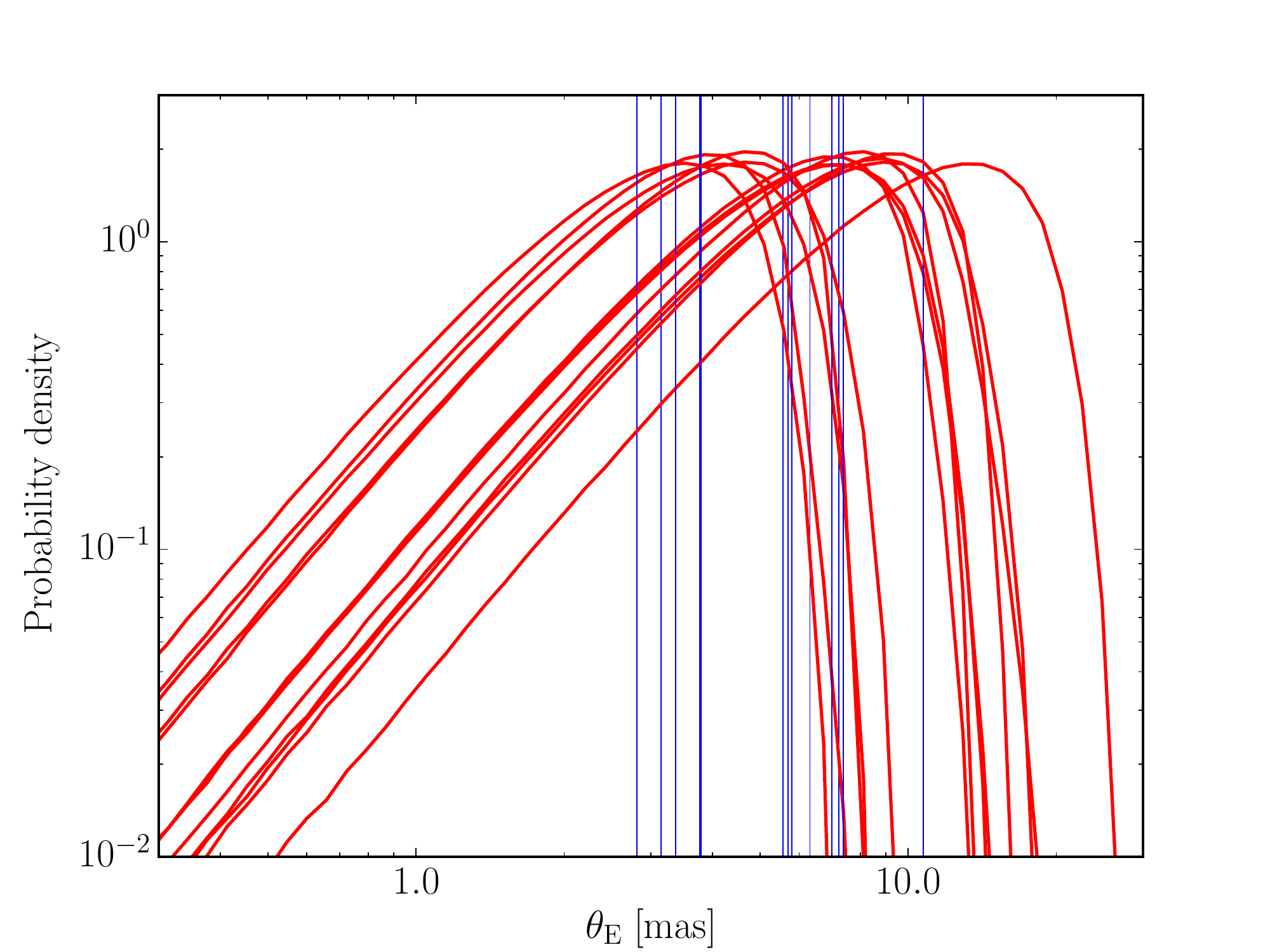} 
\caption{Distribution of probability for angular Einstein radius ($\thetaE$ ) for 13 parallax microlensing events with dark remnant lens. Vertical thin line indicate median, most probable value.  Most events with $\thetaE>1$ mas will have their Einstein radii measured from Gaia astrometry.}
\label{fig:thetaE}
\end{figure}

Fig. \ref{fig:thetaE} shows distributions of probable ranges of $\thetaE$ for dark lens candidates.
They range from a fraction of mas to couple of milliarcseconds. 
According to \cite{deBruijne2014} ESA's Gaia mission at the end of the mission should achieve astrometric precision better than 1 mas for stars as faint as $V\sim19$ mag. 
Gaia is collecting observations since mid-2014 and is expected to do so until 2019/2020. 
Therefore, for recently on-going microlensing events  with heavy remnant lenses, similar to our sample, we will have superb astrometric time-series provided by Gaia. Once combined with ground-based (or space-based) parallax measurements, a mass and distance of the lens will be measured without the need of any assumptions about proper motions. 

\begin{figure*}
\centering
\includegraphics[width=5.5cm]{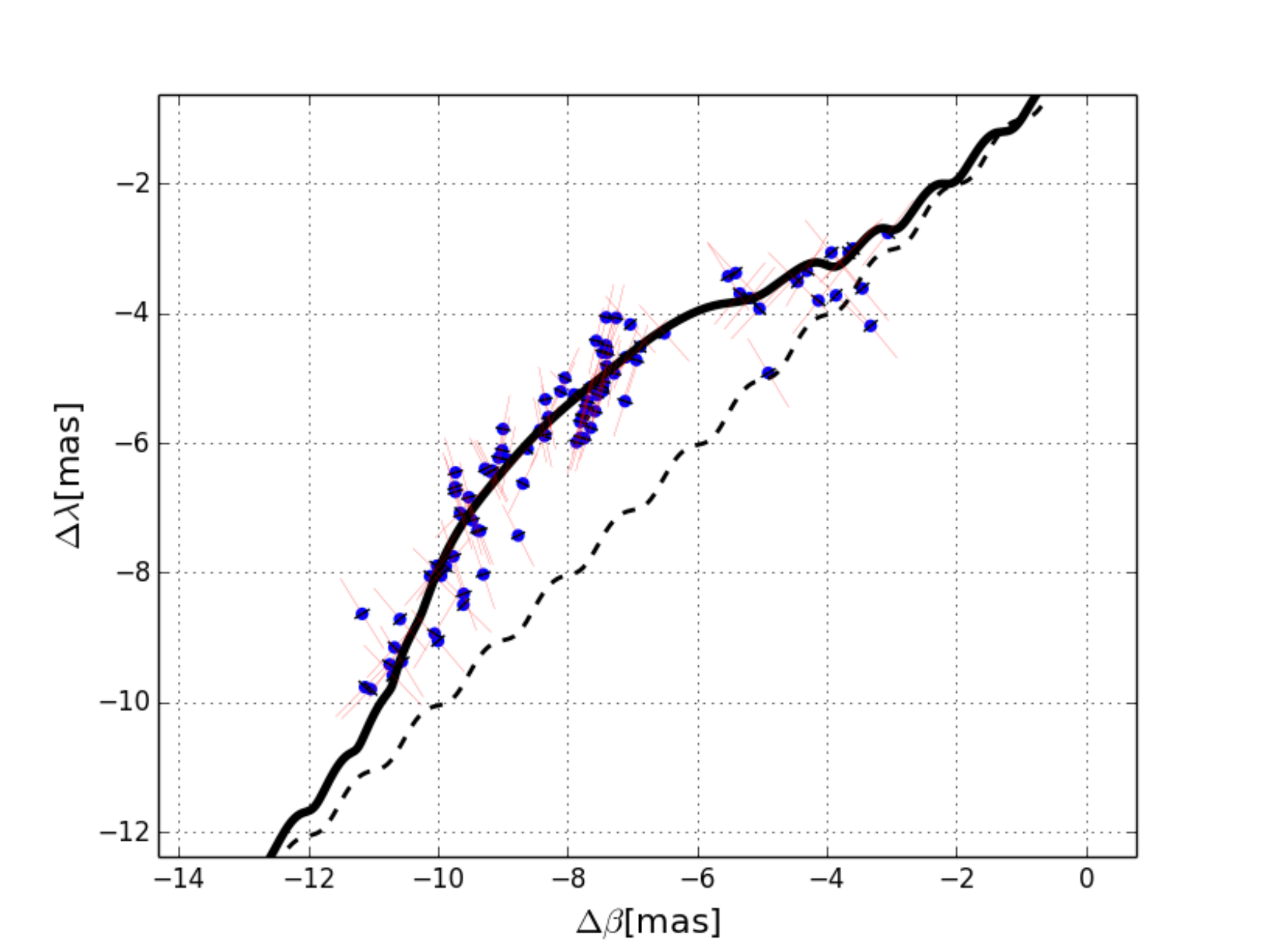}
\includegraphics[width=5.5cm]{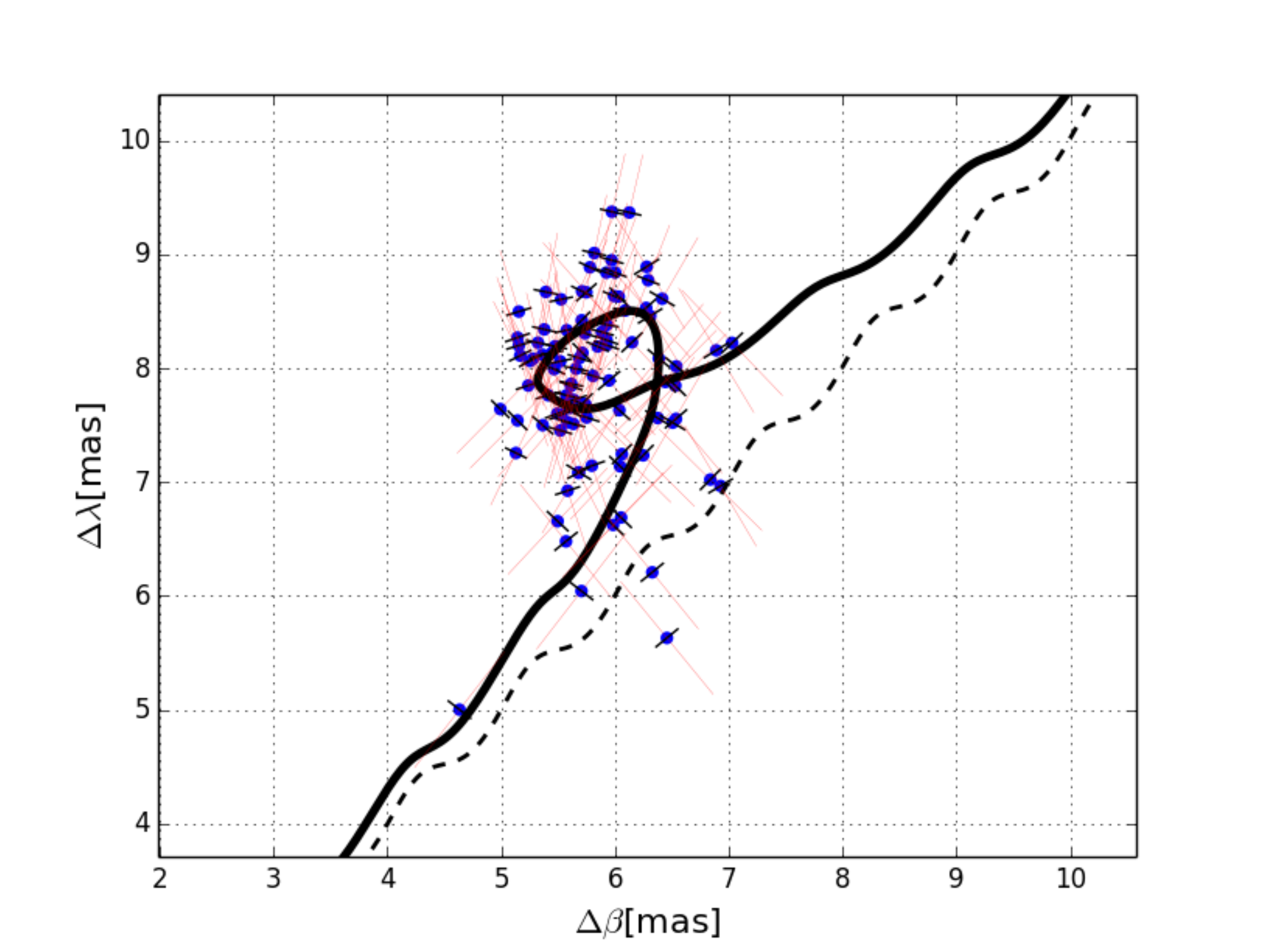}
\includegraphics[width=5.5cm]{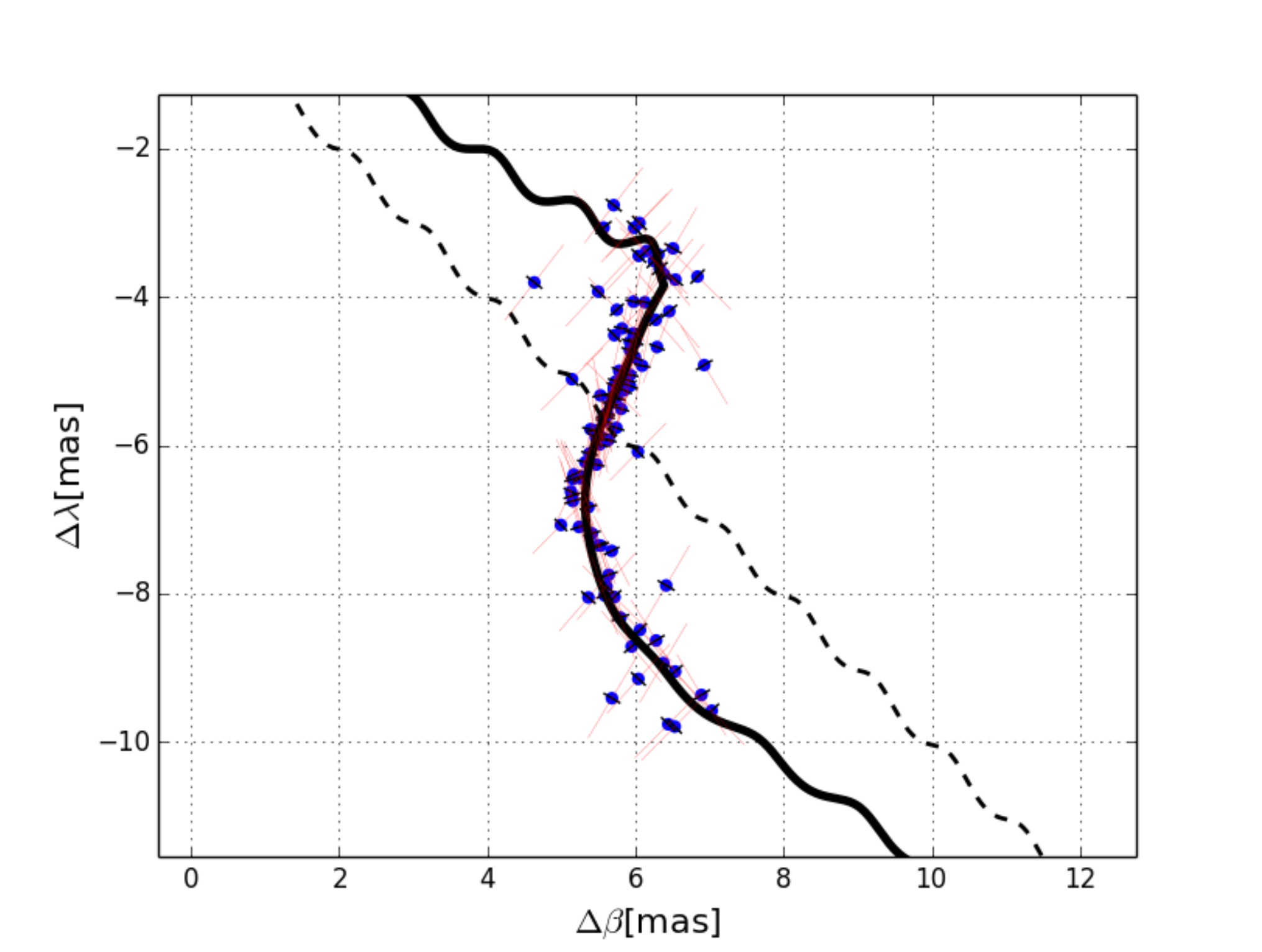}
\caption{Simulations of example astrometric paths (solid lines) for a black-hole event similar to OGLE3-ULENS-PAR-02 for $\murel=4$ mas/yr, but with different directions of that vector. Dashed lines show unperturbed source motion with parallax. Dots show mock Gaia measurements with Gaia sampling and expected accuracy. Note that error-bars along the scanning direction (blue) are significantly better than across the scanning (thin red lines).}
\label{fig:astrometry}
\end{figure*}

Several searches for those minute astrometric centroid displacements are being conducted, however, so far none were reported positive. In the near future, however, the Gaia space mission (\eg \cite{deBruijne2014}) will provide sub-milliarcsecond positional time-series for a billion stars in our Galaxy, down to $V\sim20$ mag. Such precise astrometry will be most suitable for detecting and measuring microlensing astrometric signals (\citealt{BelokurovEvans2002}, \citealt{Wyrzykowski2012}). 
Fig. \ref{fig:astrometry} shows examples of simulated astrometric trajectories for our best BH candidate, along with simulated Gaia observations, as if Gaia was launched a decade earlier. Similar events with black hole lenses will produce similar paths, which will allow for independent measurements of $\thetaE$. 
When combined with $\piE$ (obtained either from Earth orbit accelerationÕs or from space, \eg from Spitzer). Gaia will provide direct measurements of masses of lenses for hundreds of microlensing events. In particular, dark lenses would produce easy-to-measure astrometric displacements, as the astrometry would be less affected by blending. Detections of neutron stars and black holes via astrometric and photometric channels will open a new path of studying the mass function of dark remnants. 
However, we stress that it would not be possible without long-term photometric monitoring programmes. 

\section{Conclusions}
We have found 13 candidates for stellar dark remnants using microlensing events with measurable parallax effects. Among them are both neutron stars and black holes, as microlensing can not distinguish between the two, but we used the modelled light to rule out stellar luminous lenses. 
We estimated the masses and distances of those objects assuming standard proper motions for the lenses.
The most massive lens found in our sample is OGLE3-ULENS-PAR-02 with $8.7^{+8.1}_{-4.7} \msun$ for the more likely of the two solutions, and $9.3^{+8.7}_{-4.3} \msun$ for the other solution.  
Another candidate, OGLE3-ULENS-PAR-05, with most likely mass of 3.3 or 4.8 $\msun$, seems to reside in the mass gap in the mass distribution for NSs and BHs and indicates there might be a population of isolated massive remnants filling the observed gap, in line with the current theoretical predictions.

Our mass estimates rely on certain assumptions on distances and relative proper motion. 
In principle, it can not be currently fully ruled out that some of our dark remnant candidates are simply main sequence stars in the disk, which happened to co-move in parallel to a bulge source star, yielding very small relative proper motion. 
Also, we can not fully exclude a possibility that the lower mass lenses (around and below 1 $\msun$) are nearby white dwarfs. 
However, the most heavy lenses found in our study still remain the best candidates for isolated black holes within or outside the mass gap between neutron stars and black holes. 
The approximate mass function obtained for our 13 candidates resembles the one obtained in the stellar evolution synthesis and indicates no mass gap above 2 $\msun$.
The sample of microlensing events with a robust parallax measurement will be further expanded with many events being currently found in the OGLE-IV survey.

We also note that the events with the greatest length are more likely to be due to massive lenses and exhibit strong parallax effects, allowing us to constrain the mass of the lens. Conducting a search on combined OGLE Bulge data from 1997 (OGLE-II, OGLE-III and OGLE-IV) would yield discoveries of the most massive lenses and black holes ever.
On the other hand, the currently on-going OGLE-IV project is finding about 2000 events every year. Those events are also currently observed by Gaia, which at the end of its five+ years long mission will provide sub-milliarcsecond astrometric time-series for them. Combining both ground-based superb photometry and space astrometry will yield many discoveries of black hole lenses among microlensing events and will allow us to put constrains on the remnant mass function.

\section*{Acknowledgments}
The authors would like to thank Drs. Martin Dominik, Shude Mao, Vasily Belokurov, Matt Auger, Morgan Fraser, 
Tomek Bulik, Kris Belczy{\'n}ski and Virginie Batista for their continuing support to this project and Achille Nucitta, Chris Kochanek and Dave Bennett for their constructive comments.
We also would like to thank the anonymous referee for their positive attitude and comments.
This work has benefited from help of students of 2014/2015 Astronomy-III course at the Warsaw Astronomical Observatory, particularly: Ania Jacyszyn and Barbara Handzlik, and monographic lecture on Data Mining in Astronomy 2015, in particular Piotr Konorski, Piotr Wielg{\'o}rski, Paulina Karczmarek, Bartek Zgirski and Patryk Pjanka. 

{\L}. W. acknowledges support from the Polish NCN ÓHarmoniaÓ grant No. 2012/06/M/ST9/00172.
The OGLE project has received funding from the National Science Centre, Poland, grant MAESTRO 2014/14/A/ST9/00121 to AU. 
N. J. R. is a Royal Society of New Zealand Rutherford Discovery Fellow.

\label{lastpage}

\bibliographystyle{mn2e}

\end{document}